\documentclass[12pt,a4paper]{article}
\pdfoutput=1 
\usepackage{graphicx}
\usepackage{jcappub}
\usepackage{amsmath,amsthm,latexsym,amssymb,amsfonts,epsfig}
\usepackage{hyperref}
\usepackage{psfrag}
\usepackage[cp1250]{inputenc}
\usepackage[usenames,dvipsnames]{xcolor}
\usepackage{float}
\usepackage{tabularx}

\newcommand{\be}{\begin{equation}}
\newcommand{\ee}{\end{equation}}
\newcommand{\bea}{\begin{eqnarray}}
\newcommand{\eea}{\end{eqnarray}}
\numberwithin{equation}{section}
\usepackage{physics}
\numberwithin{equation}{section}

\title{Requiem to ``Proof of Inflation" or Sourced Fluctuations in a Non-Singular Bounce}

\author[a]{Ido Ben-Dayan,}
\author[a]{Udaykrishna Thattarampilly}
\affiliation{Physics Department, Ariel University, Ariel 40700, Israel}

\emailAdd{ido.bendayan@gmail.com}
\emailAdd{uday7adat@gmail.com}

\abstract{Popular wisdom suggests that measuring the tensor to scalar ratio $r$ on CMB scales is a ``proof of inflation" since one generic prediction is a scale-invariant tensor spectrum while alternatives predict $r$ that is many orders of magnitude below the sensitivity of future experiments. A bouncing Universe with sourced fluctuations allows for nearly scale-invariant spectra of both scalar and tensor perturbations challenging this point of view. Past works have analyzed the model until the bounce, under the assumption that the bounce will not change the final predictions. In this work, we discard this assumption. We explicitly follow the evolution of the Universe and fluctuations across the bounce until reheating. The evolution is stable, and the existence of the sourced fluctuations does not destroy the bounce. The bounce enhances the scalar spectrum while leaving the tensor spectrum unchanged. The enhancement depends on the duration of the bounce - a shorter bounce implies a larger enhancement. The model matches current observations and predicts any viable tensor-to-scalar ratio $r\lesssim 10^{-2}$, %which is similar to predictions of small field models of Inflation.}  
which may be observed in upcoming CMB experiments. Hence, a measurement of $r$ will no longer be a ``proof of inflation'', and a Sourced Bounce is a viable paradigm with distinct predictions.}
 
\keywords{}

\begin{document}

\maketitle
\section{Introduction}
The physical epochs predating radiation domination are key to understanding the large degree of homogeneity and isotropy of the observable Universe and the initial seeds of structure formation in the Universe. The leading paradigm of inflation suggests that a period of accelerated expansion predates radiation domination \cite{Martin:2013tda}. Extrapolating back in time before inflation, one hits the Big Bang singularity \cite{Borde:1996pt,Borde:2001nh}, which cannot be dealt with the standard effective field theory techniques\footnote{We would like to point out to the reader that these results have been under scrutiny recently \cite{Lesnefsky:2022fen} and past completeness of inflationary space-times remains an open issue.}. 
An interesting alternative is the idea of a cosmological Bounce, where there is no Big Bang singularity. The Universe contracts and at finite subplackian curvature starts expanding. The whole process is described by effective field theory (EFT). During the contraction, the Universe homogenizes and quantum fluctuations are the seeds of structure formation, similar to inflation \cite{Brandenberger:2012zb,BATTEFELD20151}.
Within the realm of EFT, a realistic bouncing scenario implies a violation of the Null Energy Condition (NEC) which could result in instabilities invalidating the analysis\footnote{In a Universe with positive spatial curvature, a bounce can be achieved by violating only the Strong Energy Condition. However, a period of inflation is then needed to dilute the curvature to the level observed today. Such a model is not an alternative to inflation, and it is unclear whether it has any distinct predictions. In general, in the paper, when we discuss alternatives to inflation that resolve the Big Bang singularity, we mean alternatives that also have definite predictions for the observed CMB spectrum.}. Therefore, careful treatment of the bounce phase is necessary to ensure a successful and realistic cosmological bouncing model, i.e. a model that is self-consistent, predictive, and in accord with current observations.

Early Universe models based on quantum fluctuations predict a scalar (density) and tensor (gravitational waves, GW) spectrum of fluctuations. These are the signatures of various early Universe models. These signatures are imprinted on the celebrated cosmic microwave background radiation (CMB), as different models result in different temperature and polarization correlations or spectra.
%The GW spectrum is sensitive only to the scale factor $a$, and therefore provides a direct probe of the geometry and expansion of the Universe.  
%As a result, a generic feature of Inflation is a nearly scale invariant GW spectrum. 
The scalar spectrum is sensitive to the field content and to the equation of state $w$. Since most inflation models imply $w\simeq-1$, a generic feature of inflation is a nearly scale-invariant scalar spectrum. 
The fact that CMB measurements do observe a nearly scale-invariant scalar spectrum is the best verification of the inflationary paradigm. 
For bouncing models, the situation is less generic. One can consider different equations of state during the contraction resulting in different predictions \cite{Lehners:2010fy,BATTEFELD20151}. For example, the simplest, and certainly most conservative, attempt for a `bouncing cosmology' is the matter bounce. In this scenario the energy density of the Universe during the bounce regime is mostly due to dust (equation of state $w$=0). A viable matter bounce model that is consistent with observations is studied in \cite{Raveendran:2017vfx, Raveendran:2018why}. However, matter bounce models tend to suffer from shear and BKL instabilities, which in turn require an ekpyrotic phase. Hence, it makes sense to discard the contracting matter dominated phase, and consider solely ekpyrotic contraction, if it can be matched to CMB observations.

A prominent example of a contracting Universe is indeed that of ekpyrotic contraction with $w\gg1$, which is free of shear and BKL instabilities  \cite{BATTEFELD20151,Khoury:2001wf}. The single field version predicts a blue scalar spectrum contrary to CMB observations. This can be remedied by the inclusion of another field - scalar or gauge field \cite{BATTEFELD20151,r3,r1,Artymowski:2020pci,r4}. 

Contrary to the scalar spectrum, which depends on the potential and the number of fields in different models, the GW spectrum directly probes the geometry of space-time, since the relevant Mukhanov-Sasaki equation only includes the scale factor $a$ and its derivatives. Hence, a scale-invariant GW spectrum that could be observed in the CMB BB polarization is a model-independent core prediction of inflation, even more robust than the prediction regarding the scalar spectrum. Better yet, for vacuum fluctuations, the GW spectrum directly probes the energy scale of inflation.
Such a measurement is specified by the tensor-to-scalar ratio $r$ and has evaded the CMB community so far, ruling out various inflation models. The most up-to-date measurements have placed an upper bound of $r<0.036$ \cite{Tristram:2021tvh,Tristram:2020wbi}.
Finally, most alternatives to inflation, and certainly the ekpyrotic model, predict a blue GW spectrum that cannot be observed in the foreseeable future without violating BBN bounds by many orders of magnitude.  Hence, a measurement of the GW signal is considered a  ``proof" of inflation. See also \cite{Brandenberger:2011eq,Chen:2018cgg,Vagnozzi:2022qmc}. 

As with any proof in Physics, there is always a loophole. The blue GW spectrum predicted in bouncing models is based on considering only vacuum fluctuations. One can consider models that have sourced fluctuations on top of the vacuum ones. The paradigm of sourced fluctuations can be realized in both inflationary and contracting backgrounds and has a rich phenomenology \cite{Chowdhury:2016aet, Chowdhury:2015cma, Chowdhury:2018blx, Gasperini:2017fqw, Ito:2016fqp, Lin:2015nda, Wang:2014kqa, Caprini:2014mja, Barnaby:2012xt,r1,r3}. The sourced fluctuations are a result of a coupling between the scalar field driving the evolution of the Universe and other fields whose energy density is subdominant compared to the scalar field but are large enough to generate a source term in the equations for fluctuations. As a result, two types of spectra are generated - vacuum one and sourced one with no cross terms\footnote{ In terms of Feynman diagrams, if the vacuum spectrum is the tree level 2-point function, the sourced spectrum is a 1-loop calculation.}.
Most importantly, we have shown that considering sourced fluctuations can result in a nearly scale invariant chiral GW spectrum in bouncing models discarding the proof of inflation \cite{r3,r1}. 

Our previous works realized this idea by considering a $U(1)$ gauge field coupled to the scalar field that drives the contraction, the ekpyrotic field. We have analyzed various possible couplings between the fields and various potentials. The requirement of a nearly scale-invariant scalar spectrum that matches the CMB allowed only specific forms of couplings. As a result, the GW prediction was $r\simeq \frac{1}{9}$, which is above the aforementioned upper bound \cite{Artymowski:2020pci}.

These results were analyzed using a canonical kinetic term for the scalar field. A canonical scalar field does not violate the NEC, so it does not show how the universe and spectra evolve through the bounce. Hence, our findings were based on the assumption that the bounce will not alter the conclusions.  In this work, we insert the final piece into the puzzle. We consider the full action, which includes non-canonical kinetic terms of the scalar field and takes into account the bounce phase. Such a class of models was first studied in the context of dark energy and late time acceleration in  \cite{Deffayet:2010qz}. Fields with non-canonical kinetic have been known to be able to produce non-singular bounce 
 \cite{Cai:2012va,Qiu:2015nha,Easson:2011zy,Koehn:2015vvy} without instabilities. Observational signatures of such a bounce followed by inflation have been studied previously in \cite{Ni:2017jxw}.

We consider a model of non-singular bounce prescribed in \cite{Cai:2012va} as the background model. The model involves Galileons with a non-canonical kinetic term. We couple the Galileon field to a $U(1)$ gauge field as has been done in previous works for the canonical scalar field. Since Galileon has a non-canonical kinetic term, its kinetic energy can temporarily be negative. When the kinetic energy becomes negative, the scalar field enters a regime of ghost condensation, kick-starting the bouncing mechanism. 
We evolve the universe through the bounce all the way up to reheating and radiation domination. We show that the evolution is stable, even with the inclusion of gauge fields, given that the energy density of gauge fields remains small enough. 

We then evolve the vacuum and sourced scalar and GW spectra.  %In order to analytically solve the perturbation equations, we modify the background relations for the scalar field in the context of our numerical results.
We solve the perturbation equations both analytically and numerically. Perturbations in Galileon theories could be plagued by ghost and gradient instabilities \cite{Libanov:2016kfc}. Methods to circumvent instabilities in non-singular bounce and bounce-inflation theories are discussed in the literature \cite{Cai:2016thi,Cai:2017tku,Cai:2017dyi}. The short duration of the bounce renders the growth of perturbations due to gradient instabilities under control. % Within our analytic approximation, all scalar modes are amplified by the same finite factor that depends on the duration of the bounce. Tensor modes are unchanged, and perturbative control is maintained throughout the calculation.
%We also numerically verify that the evolution of sourced perturbations remains void of instabilities.}      
Our analysis shows that the scale dependence of the spectra is unaffected by the Bounce. The GW modes are practically unchanged. All scalar scalar modes are amplified by the same finite factor that depends on the duration of the bounce, while staying in the perturbative regime throughout the calculation. We also numerically verify that the evolution of sourced perturbations remains void of instabilities.
As a result, we have a viable model with the following predictions for the scalar spectrum amplitude and tilt, $A_s=2.1\times 10^{-9},\, n_s\simeq 0.96$  and the tensor to scalar ratio
\be
r\lesssim10^{-2}
\ee
within the observable reach of SO or CMB-S4 \cite{CMB-S4:2020lpa,Hensley:2021ydb,Namikawa:2021gyh}. 
%\cite{LIGOScientific:2007gwp,LIGOScientific:2006zmq,Fotopoulos:2008yq,LIGOScientific:2009qal,LIGOScientific:2019vkc,}.
Moreover, our model predicts GW with one chirality. Thus, if one can measure $r$ to such sensitivity, then we predict the GW will be chiral, contrary to inflationary predictions. The model can further be tested by analyzing the predictions for Laser Interferometers, as well as measurements of the chirality of the spectrum \cite{LIGOScientific:2019vkc,Abbott:2016blz,AmaroSeoane:2012km,PhysRevD.81.123529}. We defer the analysis of signals in Laser Interferometers to future work. 

In brief, we have a viable alternative paradigm of "Sourced Bounce" with distinct predictions for Early Universe physics. The paradigm suggests that the Universe contracts and then expands without a singularity in a stable self-consistent analysis. The observed CMB spectra are due to sourced fluctuations. More generally, within this paradigm, we can construct various models with different scalar potentials, different sources and different couplings and derive predictions for the Early Universe. Measuring $r$ will not be proof of inflation, and corroborating any paradigm of the Early Universe will require further measurements such as non-gaussianity, the tensor tilt $n_T$, chirality and GW on other scales.
%Finally, as discussed the GW vacuum spectrum of bouncing models is blue, which immediately raises the question of implications by LIGO and LISA. We defer the analysis of this issue to future work.

%We wish to stress that the "Sourced Bounce" is a paradigm and not a model since one can consider various other sources such as $SU(2)$ gauge fields or modifications of the kinetic term will change the prediction of $r$ or of the scalar spectrum.

The paper is organized as follows. In section \ref{sec:nonsinb} 
 we discuss our setup. One part is the well-established non-singular bounce
modeled by a single non-canonical scalar field. On top of that, we add a U(1) gauge field coupled to the Galileon. In section \ref{sec:vacev}, we evolve the vacuum perturbations of our model through the non-singular bounce. 
Section \ref{sec:sourced} discusses the general method we employ to evaluate perturbations sourced by a gauge field and their evolution across the bounce. Tensor perturbations and their evolution are discussed in detail here. The following section, \ref{sec:scalar}, discusses the evolution of sourced scalar fluctuations by deriving perturbation equations for sourced scalar perturbations from third-order action and solving them. We present both analytical and numerical solutions. Section \ref{sec:fr} discusses the era of kinetic domination after the bounce leading up to the standard radiation, matter and dark energy-dominated eras of standard cosmology. We then conclude.

\section{Setup}
\label{sec:nonsinb}
\subsection{The background model}
In previous works, we have examined the evolution of perturbations sourced by a $U(1)$ gauge field and calculated tensor and scalar perturbations produced during the ekpyrotic phase \cite{r1,r4,Artymowski:2018ewb}. We deduced that the tensor-to-scalar ratio is larger than the observed bounds measured by recent experiments \cite{Paoletti:2022anb}. However, this result rests under the assumption that the perturbations are unchanged as they cross the bouncing phase of the Universe. In this article, we explicitly evaluate the change in the power spectrum of tensor and scalar perturbations as the Universe undergoes a non-singular bounce. The non-singular bouncing Universe that we are examining is modeled by the Lagrangian \cite{Cai:2012va},
\begin{equation}
\mathcal{L} = K(\phi,X)+G(\phi,X)\Box \phi,
\label{eq:sclag}
\end{equation}
where 
\begin{equation}
 K(\phi,X) = (1-g(\phi)) X+ \beta X^2-V(\phi)
 \label{eq:kin}
\end{equation}
\begin{equation}
 G(\phi,X) = \gamma X.
\end{equation}
Here, $X=-\frac{1}{2}\partial_{\mu}\phi \partial^{\mu} \phi$. The simplest form of  $K(\phi,X)$ for models involving a non-canonical field is $(1-g(\phi)) X-V(\phi)$, however, such a function cannot produce a non-singular bounce. In order to obtain a successful bounce, we add an additional quadratic term to $K(\phi,X)$.
Upon choosing the potential to be
\begin{equation}
V(\phi) = -\frac{2V_0}{e^{-\sqrt{\frac{2}{q}}\phi}+e^{b_V\sqrt{\frac{2}{q}}\phi}}
\end{equation}
and the modification of the kinetic term 
\begin{equation}
g(\phi) = \frac{2g_0}{e^{-\sqrt{\frac{2}{p}}\phi}+e^{b_g\sqrt{\frac{2}{p}}\phi}} \, ,
\end{equation}
one will obtain a phase of ekpyrotic contraction away from the bounce. This model gives rise to a successful non-singular bounce as explained in \cite{Cai:2012va}. The vanishing of the covariant derivative of the stress-energy tensor will give the generalized Klein-Gordon equations for the background.
\begin{figure}[H]
    \centering
    \includegraphics[width=0.49\textwidth]{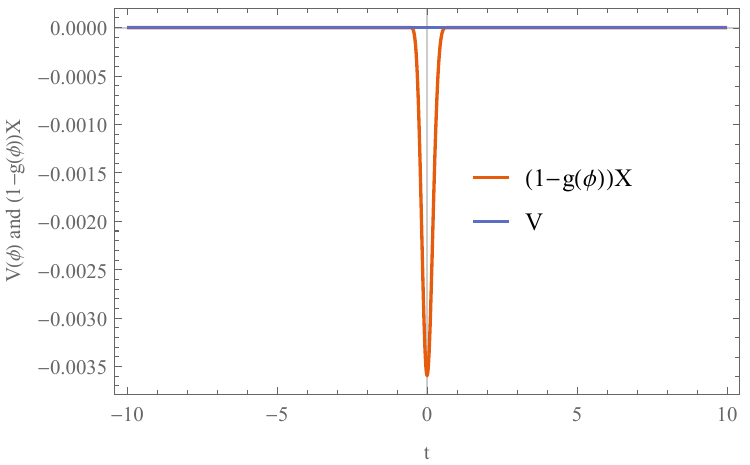} 
    \hfill
    \includegraphics[width=0.49\textwidth]{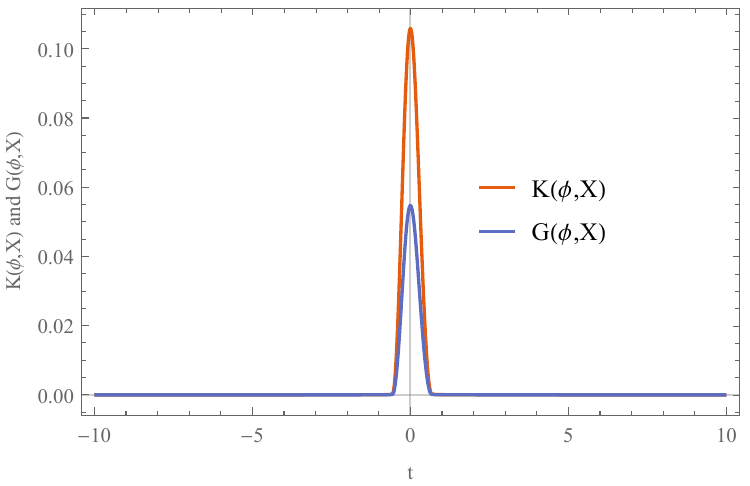} 
     \caption{ Comparison of the potential and kinetic term (left) and functions $K(\phi,X)$ and $G(\phi,X)$ (right) as a function of time $t$, near the bounce. $G(\phi,X)$ is rescaled to the order of $K(\phi,X)$ by multiplication of a constant real number.} 
\label{fig:eqxv}
\end{figure}
In figure \ref{fig:eqxv}, we show an example of the potential and kinetic terms, and the functions $K(\phi,X)$ and $G(\phi,X)$. Near the bounce, the kinetic terms dominate over the potential, and non-canonical effects become significant. 
On the other hand, $g$ is close to zero for large values of the field. Therefore, the Lagrangian away from the bouncing phase can be approximated as a standard canonical scalar field with a potential
\begin{equation}
    \mathcal{L} = -\frac{1}{2} \partial_{\mu} \phi \partial^{\mu} \phi -V(\phi),
\end{equation}
which will admit ekpyrotic solution $w_c \gg1 \Leftrightarrow q \ll 1$ with equation of state
\begin{equation}
    w_c = -1+ \frac{2}{3q}. 
    \label{eq:ekpw}
    \end{equation}
 
\begin{figure}[H]
    \centering
    \includegraphics[width=0.8\textwidth]{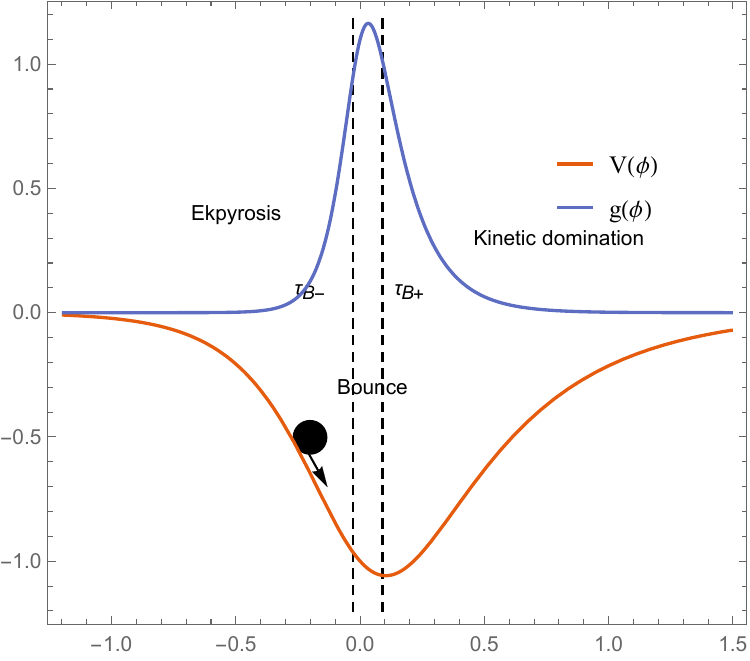} 
     \caption{Example of the potential of the scalar field $V(\phi)$ and the kinetic function $g(\phi)$. Vertical lines represent the beginning and end of the ghost condensation phase.} 
\label{fig:vg}
\end{figure}

\begin{figure}[H]
    \centering
    \includegraphics[width=0.49\textwidth]{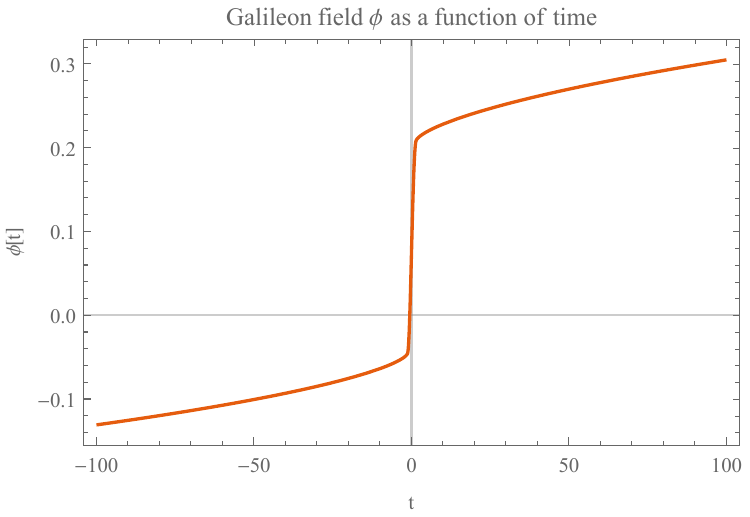} 
    \hfill
    \includegraphics[width=0.49\textwidth]{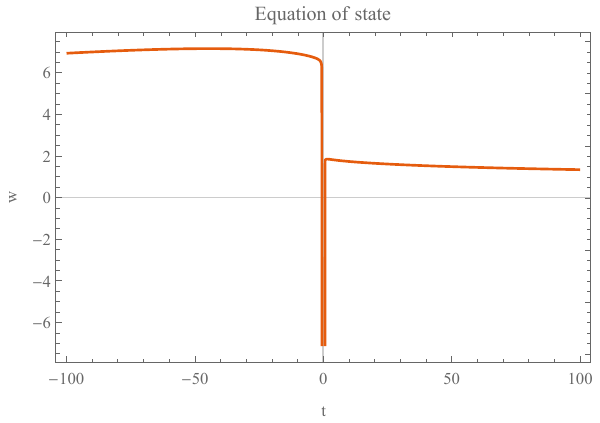} 
     \caption{ Galileon field $\phi(t)$ and equation of state of $\omega_{\phi}(t)$ as a function of cosmic time $t$.} 
\label{fig:eqnst}
\end{figure}

Figure (\ref{fig:vg}) shows a graph of the scalar field potential $V(\phi)$ and the non-canonical term $g(\phi)$ as a function of the scalar field $\phi$. $\phi_{B-}(\tau_{B-})$ and $\phi_{B+}(\tau_{B+})$ represent the beginning and end of the bouncing phase (and NEC violation).
The Universe starts at $\phi \ll -1$ with a slow ekpyrotic contraction. As $\phi$ accelerates towards $\phi$ = 0, the value of g will
increase. If $g(0)> 1 $  (which we require), then at some point
in time, g will exceed the critical value of $g = 1$
and the sign of the kinetic term $X$ in \eqref{eq:kin} will
become negative, giving rise to a phase of ghost condensate. This phase of ghost condensation coincides with the bouncing phase of the Universe. This is the region in field space where the NEC is violated, which in turn triggers the bounce at $\phi=0$. 
The Universe continues to roll to positive larger values of the field $\phi>\phi_{B+}$, after which the Universe enters the era of kinetic energy domination. %Here, the short-lived bounce phase is followed by an era of kinetic expansion. 
The Lagrangian recovers its canonical form after $\phi_{B+}$. Notice that despite the approximate symmetry of the potential, the scalar field $\phi$ does not approach an ekpyrotic solution, unlike the contracting phase of $\phi \ll -1$, since this solution in an expanding Universe is not an attractor. Instead, the field $\phi$ goes through a fast rolling phase with an equation of state $w=1$. This fast roll phase results in a dilution of the energy density of scalar field $\phi$, leading up to a radiation-dominated era of the Universe. 
Finally, figure \ref{fig:eqnst} shows the field value and the equation of state of the Galileon field as a function of time. Clearly, the equation of state $w\gg1$ before the bounce and asymptotes to one after the bounce.
%\section{Non-singular bouncing Universe coupled to a $U(1)$ gauge field}
\subsection{Adding gauge fields and the EOM of the scalar field}
We want to model a non-singular bouncing Universe where the perturbations are sourced by a $U(1)$ gauge field, similar to the contracting Universe models in \cite{r1,r4,Artymowski:2018ewb}. Hence, we add a $U(1)$ gauge field and couple it to the scalar field driving the contraction in \eqref{eq:sclag}. %The scalar field in the equation  will now be coupled to a $U(1)$ gauge field.  
This model is described by the full action: 
\begin{equation}
S = \int d^4x \sqrt{-g} \left[\frac{R}{2}-K(\phi,X)+G(\phi,X)\Box \phi-I^2(\phi) \left( \frac{1}{4}F^{\mu\nu}F_{\mu \nu} -\frac{\delta}{4}\tilde{F}^{\mu \nu}F_{\mu\nu}\right) \right] \label{eq:vecaction}
\end{equation}
where $\phi$ is the bouncer field, $A_{\mu}$ is the U(1) gauge potential,  $F_{\mu\nu} = \partial_\mu A_\nu - \partial_\nu A_\mu$, $\tilde{F}^{\mu\nu} = \frac{1}{2}\epsilon^{\mu\nu\rho\sigma}F_{\rho\sigma}$,  $\delta>0$ is a coupling constant. 
Using the flat FLRW metric in cosmic time $t$ or conformal time $\tau$,
\be
ds^2=-dt^2+a^2(t) d\vec{l}^2=a^2(\tau)[-d\tau^2+d\vec{l}^2],
\ee
equations of motion are derived from the Einstein field equations given by $(M_{pl}^{-1}=\sqrt{8\pi G_N}=1)$:
\begin{equation}
    G_{\mu \nu} = T_{\mu \nu}.
\end{equation}
For our theory, including the gauge fields 
\begin{equation}
\begin{split}
    T_{\mu \nu} &= g_{\mu \nu}\left(-K(\phi,X)+2 X G,_{\phi} + G,_{X} \grad_{\mu} X \grad^{\mu} \phi \right) \\
    & + (K,_X- 2 G,_{\phi}+G,_X \Box \phi) \grad_{\mu} \phi \grad_{\nu} \phi \\
    &- G,_{X} (\grad_{\nu} X \grad_{\mu} \phi+\grad_{\mu}X \grad_{\nu} \phi ) \\
    & + I^2(\phi) \left(F_{\mu}^{\rho} F_{\nu\rho} - \delta \tilde{F}_{\mu}^{\rho} F_{\nu\rho} \right) - \frac{I^2(\phi)}{4} g_{\mu \nu}\left(F^{\rho \sigma}F_{\rho \sigma}-\delta F \Tilde{F} \right).
    \end{split}
    \label{eq:emt}
\end{equation}

If the energy density stored in the gauge fields is negligible compared to the bouncer field, the background evolution will be practically unaffected, and we can treat the gauge field as a higher-order perturbation. This requirement induces constraints on the parameters of the model, see below. %The addition of gauge fields does not alter the equations of motion for the scalar field since the gauge fields contribute to the equations only at higher orders in perturbation theory. 
Thus, the background evolution of the scalar field is consistent with the evolution described in  \cite{Cai:2012va}.
The equations of motion for the background field are of the form 
\begin{equation}
    \ddot{\phi} \mathcal{P} + \mathcal{D} \dot{\phi} + V,_{\phi} = 0,
    \label{eq:eom}
\end{equation}
where
\begin{equation}
  \mathcal{P} = (1-g) + 6 \gamma H \dot{\phi} + 3 \beta \dot{\phi}^2 + \frac{3 \gamma^2}{2} \dot{\phi}^4
  \label{eq:p}
\end{equation}
and
\begin{equation}
\begin{split}
    \mathcal{D}  = 3 (1-g) H +(9 \gamma H^2-\frac{1}{2} g,_{\phi}) \dot{\phi} + 3 \beta H \dot{\phi}^2 -\frac{3}{2} (1-g) \gamma \dot{\phi}^3 -\frac{9 \gamma^2 H \dot{\phi}^2}{2} -\frac{3 \beta \gamma \dot{\phi}^4 }{2}. 
    \end{split}
    \label{eq:d}
\end{equation}
Denoting $\rho_g$ as the energy density of gauge fields, the first Friedmann equation is then %For this model, the background energy density is 
\begin{equation}
    3H^2 = \frac{1}{2} (1-g) \dot{\phi}^2+ \frac{3}{4} \beta \dot{\phi}^4+ 3 \gamma H \dot{\phi}^3+V(\phi)+\rho_g,
\end{equation}
%where $\rho_g$ is the energy density of gauge fields. %Friedmann equations necessitate that $3H^2=\rho$, 
which, combined with equation \eqref{eq:eom}, determine the background evolution of the universe.
\subsection{Non-Singular bounce with a gauge field}
When adding the gauge field to the equations of motion, one needs to make sure that the gauge field energy density remains subdominant compared to the energy density of the background field so as to avoid backreaction and to make sure that the non-singular bounce is undisturbed by the addition of the gauge fields. 
Following Ref. \cite{Caprini:2014mja,r1,r3} we define $\tilde{A}=IA$, where
$A$ is the gauge field, and $I$ is the coupling of the scalar field to the gauge field. $I$ is given by
\begin{equation}
    I(\phi) = \frac{1}{1+e^{-a_1 n (\phi-\phi_{B-})}}
    \label{eq:I}
\end{equation}
where $a_1 =\sqrt{ \frac{1}{2q}}$ and $\phi_{B-}$ is value of field $\phi$ at the beginning of bounce. We define a parameter $\xi= 2\pi \delta $.
$I(\phi)$ is defined such that during the regime of ekpyrosis, i.e. for large and negative $\phi$, 
\begin{equation}
     I(\phi) \simeq e^{a_1 n \phi}
\end{equation}
as defined in previous works. However for values of  $\phi>\phi_{B-}$, $I(\phi)$ is a constant. This includes the regime of bounce and kinetic domination. The constant nature of $I$ ensures that the gauge field produced during the regime of bounce and kinetic domination act as if they are in the Minkowski vacuum.\\
\begin{figure}[H]
    \centering
    \includegraphics[width=0.49\textwidth]{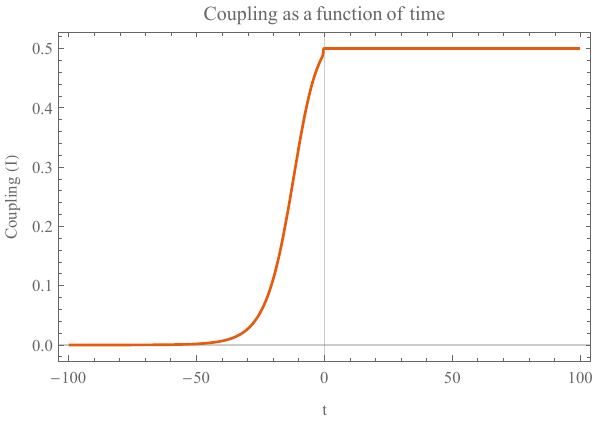} 
    \hfill
    \includegraphics[width=0.49\textwidth]{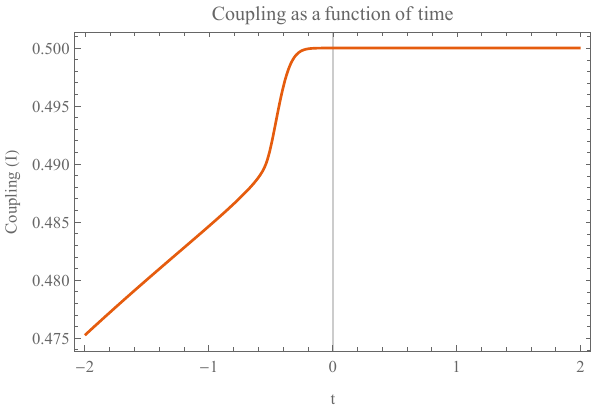} 
     \caption{Time evolution of coupling $I(\phi(t))$ of field $\phi$ to gauge fields. The right panel is zoomed in version near the bounce. For smaller values of q coupling $I$ becomes a constant faster.} 
\label{fig:gag}
\end{figure}
$\tilde{A}$ satisfies the equation of motion
\begin{equation}
\tilde{A''}_{\lambda}+\left(k^2+ 2\xi \lambda k \frac{I'}{I} -  \frac{I''}{I}\right) \tilde{A}_{\lambda} = 0 \, \label{eq:guage},
\end{equation} 
where $\lambda$ is a chiral polarization of the gauge field, and $'$ denotes differentiation with respect to conformal time $\tau$. The energy density of gauge fields as a function of time will tell us whether they remain subdominant throughout the evolution of the universe.\\

\textbf{Ekpyrotic phase $(\phi \ll -\phi_{B-},\,\tau\ll \tau_{B-})$:} During the ekpyrotic phase, %the coupling $I$ retains its exponential form as a function of $\phi$ and %coupling of the form given in \eqref{eq:I} can be rewritten in conformal time as 
$I(\tau)\simeq (-\tau)^{n}$ and $n=2$ or $n=-1$ leads to scale-invariant sourced perturbations \cite{r3,r1,Artymowski:2020pci}. The energy density of the gauge fields is therefore,
\begin{equation}
     \frac{1}{2}\bigg \langle \abs{\Vec{\tilde{E}}}^2+ \abs{\Vec{\tilde{B}}^2} \bigg\rangle \approx D(n)\frac{ e^{2\pi \xi} }{\xi^3 a^4 \tau^4} \, , \label{eq:backreactionbound}
 \end{equation}
 and places constraints on the parameters of the model, such as $\xi$. This behaviour is seemingly
 singular as we approach the bounce. \\
%   We assume that the coupling is $I(\phi)= e^{a_1 n \phi}$ throughout the evolution of the Universe. 

\begin{figure}[H] 
 \centering
        \includegraphics[width=0.45\linewidth]{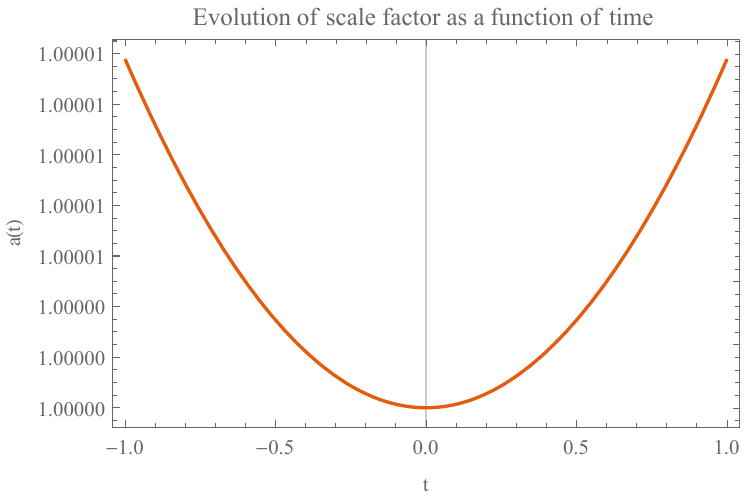} 
    \hfill
        \includegraphics[width=0.45\linewidth]{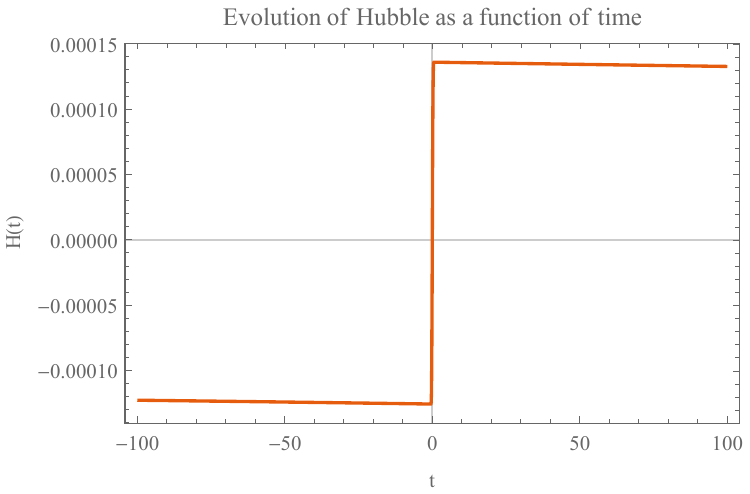}
 \vspace{1cm}
        \includegraphics[width=0.45\linewidth]{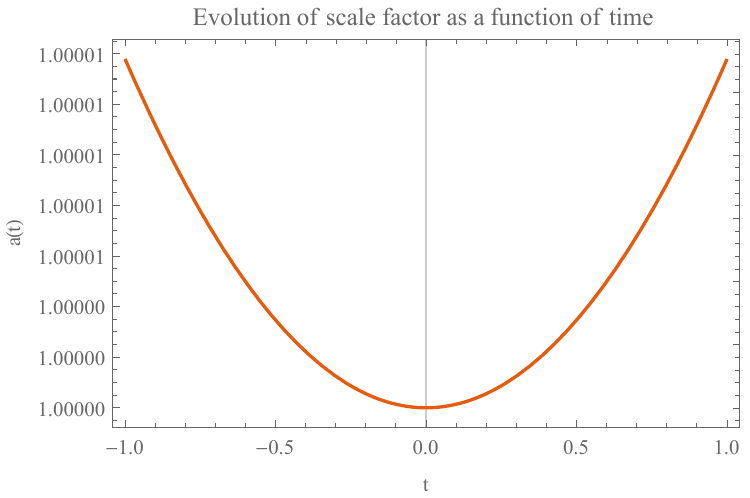} 
    \hfill
        \includegraphics[width=0.45\linewidth]{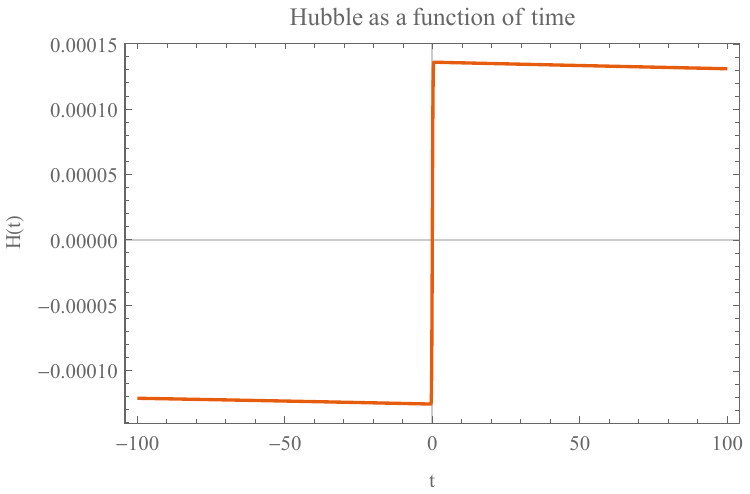} 
\caption{Evolution of the scale factor $a(t)$ (left) and the Hubble parameter $H(t)$ (right) across the bounce without gauge fields (top) and with gauge fields (bottom) given the parameters of the model specified in the text. The energy density stored in gauge fields is at least an order of magnitude smaller than the potential of the scalar field. Note that the effect of the gauge fields on the background evolution of the universe is practically indistinguishable.} 
\label{fig:background}
\end{figure}

\textbf{Bounce phase $(\phi\sim 0,\, \tau\sim 0)$:} Near the bounce $I$ will approach a constant value as shown in figure \ref{fig:gag}. For smaller values of $q$, $I$ approaches a constant value faster. Since ekpyrosis demands a small value of $q$, It is possible to approximate $I$ as a constant during the bounce phase and $I$ clearly remains a constant for positive values of $\phi$ corresponding to kinetic domination. In \cite{Cai:2012va} it is mentioned that $\dot{\phi}=\dot{\phi_B} e^{-t^2/T^2}$ near the bounce. We will modify this relation later in the article, but it is sufficient for current purposes. Near the bounce, up to first order in $t\sim \tau$ (since $a\sim 1$ near bounce), one has $\phi \approx \dot{\phi_B} \tau$. Thus, near the bounce $I(\phi)=\frac{1}{1+e^{-a_1 n (\phi-\phi_{B-})}} \sim 1- 1/\left(\left(a_1 \phi\right)^{n}-\left(a_1 \phi_{B-}\right)^n \right)\sim 1-1/\left(a_1^n (\dot{\phi}_B \tau)^{n}-\left(a_1 \phi_{B-}\right)^n\right)$. For $q\ll1$, $a_1\ll1$, given that $a_1 > \frac{1}{\abs{\phi_{B-}}}$, As $\tau \rightarrow 0$, $\phi \rightarrow 0$ and $I$ approach a constant value\footnote{ Here we are implicitly assuming $n>0$, for the $n<0$ case we have to use the duality discussed in \cite{r3}, and use $n\rightarrow -1-n$.}. Hence, towards the bounce, the gauge fields act as if they are decoupled from the scalar field. The gauge field energy density will grow and attain a constant maximum during the bounce. By requiring that this maximum is much smaller than the energy density of the background scalar field, we can keep the backreaction negligible and the bounce non-singular. In brief, the given $I(\phi)$ equations are not exactly solvable and have to be solved numerically, but the energy density during the bounce can be approximated as \cite{Caprini:2014mja,r1} 
 \begin{equation}
    \frac{1}{2}\bigg \langle \abs{\Vec{\tilde{E}}^2}^2+ \abs{\Vec{\tilde{B}}^2} \bigg\rangle \simeq D(n)\frac{ e^{2\pi \xi} }{\xi^3 a^4 \left(\tau_{B-}\right)^4}.
 \end{equation}
We numerically solve the Friedmann equations and verify that by a suitable choice of parameters, the gauge field energy density is much smaller than the energy density of the background scalar field and the bounce is non-singular. Figure \ref{fig:background} is an example solution to Friedmann equations both in the presence and the absence of gauge fields. The background evolution of the Universe near the bounce depicted in the figure corresponds with the following parameter values, $V_0 = 10^{-7},g_0 = 1.1,\beta = 5,\gamma = 10^{-3},b_V = 5,
b_g = 0.5, p = 0.05, q = 0.1, \delta=3$. The maximal energy density of gauge fields is at the bounce $t=0$, and reaches its maximal value of $10^{-1}V_0$. After the bounce, as $\phi$ grows, $I$ will remain constant. The gauge fields in \eqref{eq:guage} will simply stay in their Minkowski vacuum with no further observable effects. They will continue to redshift like radiation.
 \section{Evolution of perturbations across the Non-Singular bounce}
\label{sec:vacev}
In the absence of gauge fields acting as sourced terms, the evolution of perturbations across a non-singular bounce was examined in \cite{Cai:2012va}. Upon the addition of gauge fields, the perturbation equations transform into inhomogeneous differential equations. The solution to perturbation equations is a linear combination of homogeneous and inhomogeneous solutions, namely vacuum and sourced perturbations. In this section, we describe the known solution to perturbation equations in the absence of gauge fields. Solutions obtained here will constitute the vacuum perturbations in the presence of gauge fields acting as sources. 

\begin{figure}[H]
    \centering
    \includegraphics[width=0.48\linewidth]{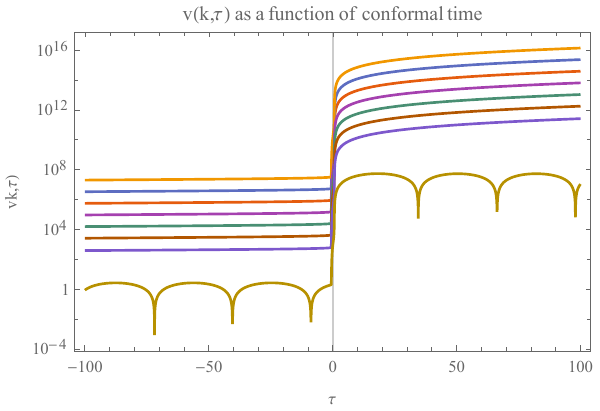} 
    \hfill
     \includegraphics[width=0.48\linewidth]{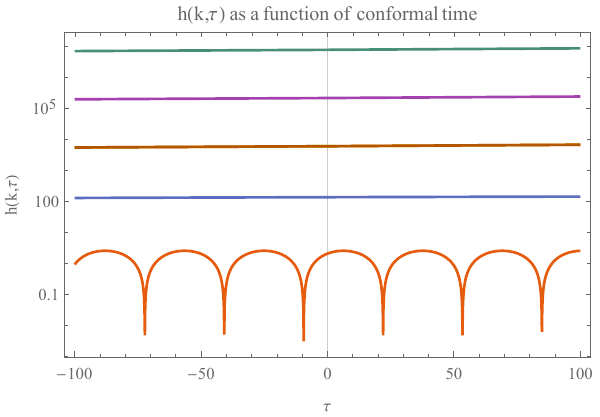} 
     \caption{Evolution of scalar (left) and tensor (right) modes for various wave numbers across the bounce as a function of conformal time. The upper curvess are modes that exit the horizon, while the bottom one is a mode that always remains subhorizon. Tensor modes are unchanged, as expected, while the scalar modes are amplified without any change in the spectral index.}
\label{fig:vacc spectrum}
\end{figure}

Equations governing scalar and tensor perturbations are obtained by expanding the action up to the second order. The perturbation theory for actions involving Galileons is well established. The Mukhanov-Sasaki variable $z$  is modified as \cite{Cai:2012va}\footnote{We have used a definition inline with the standard definition of z, $z^2=2a^2 \epsilon$ instead of $a^2 \epsilon$ as used in \cite{Cai:2012va}.}
\begin{equation}
z^2 = \frac{4a^2\dot{\phi}^2 \left((1-g)+6\gamma H \dot{\phi}+3\beta \dot{\phi}^2+\frac{3\gamma^2\dot{\phi}^4}{2}\right)}{(2H-\dot{\phi}^3 \gamma)^2}
\label{eq:bz}
\end{equation}
and the speed of sound \cite{Cai:2012va}
\begin{equation}
   c_s^2 =  \frac{ \left((1-g)+4\gamma H \dot{\phi}+\beta \dot{\phi}^2-\frac{\gamma^2\dot{\phi}^4}{2}+2\gamma \Ddot{\phi}\right)}{\left((1-g)+6\gamma H \dot{\phi}+3\beta \dot{\phi}^2+\frac{3\gamma^2\dot{\phi}^4}{2}\right)}.
   \label{eq:bcs}
\end{equation}
The equation for modes of scalar and tensor perturbations is given by
\begin{equation}
    v''(k,\tau)+\left(c_s^2 k^2-\frac{z''}{z} \right) v(k,\tau) = 0 \, ,
    \label{eq:muksass}
\end{equation}
and
\begin{equation}
    h''(k,\tau)+\left( k^2-\frac{a''}{a} \right) h(k,\tau) = 0 \, .
    \label{eq:muksast}
\end{equation}
Perturbation equations are solved approximately in various regimes. The spectrum is generated during a period of slow contraction (ekpyrotic phase). We can then deduce an approximate expression for the change in the power spectrum as the universe undergoes a bouncing phase. The vacuum tensor spectrum remains unchanged across the non-singular bouncing phase, while the scalar spectrum is amplified significantly. We have verified these results numerically. A numerical example of the evolution of the spectrum across the bounce is given in Figure \ref{fig:vacc spectrum}. 
As we mentioned in section \ref{sec:nonsinb}, the Universe undergoes an Ekpyrotic phase, a bouncing phase, and a phase of kinetic-driven expansion before reaching the standard radiation, matter, and Dark Energy dominated eras. To gain an analytical understanding of the perturbations, we solve the perturbation equations during each of these epochs and match them on the transition hypersurface between the phases. In order to continue the solutions during the transition, we make use of matching conditions derived in \cite{Hwang:1991an,Deruelle:1995kd}. The analysis matches the numerical findings. \\
\subsection{Ekpyrotic phase}
During the ekpyrotic phase characterized by an equation of state as given in equation \eqref{eq:ekpw}
\begin{equation}
    z^2 \simeq \frac{2 a^2}{q} \;\;,\;\;  a \simeq (\tau_{B-}-\tau )^{\frac{q}{1-q}} \;\;,\;\; c_s^2 \simeq 1
\end{equation}
solutions to equations \eqref{eq:muksass} and \eqref{eq:muksast} are given by the Hankel functions:
\begin{equation}
    c_1(k) H_{\frac{1}{2}+b}^{(1)}(k( \tau-\tau_{B-})) +  c_2(k) H_{\frac{1}{2}+b}^{(2)}(k (\tau-\tau_{B-}))
\end{equation}
with $b\simeq\frac{q}{(1-q)}\simeq q$.
\subsection{Bouncing phase}
When $\phi$ evolves into the ghost condensate range, the universe will exit from ekpyrotic contraction and enter a bouncing phase. During the bounce phase, the deviation of the equation for fluctuations from the canonical one becomes important. When studying fluctuations across the bouncing phase, it is convenient  to model the evolution of the Hubble parameter near the bounce as a linear function of cosmic time, i.e.
\begin{equation}
    H = \Upsilon t,
    \label{eq:hbounce}
\end{equation}
where $\Upsilon$ is a positive constant depending on the details of the bounce. Approximating the evolution of $\dot{\phi}$ as a Gaussian and employing WKB approximations, we can derive the approximate functional form of scalar modes during the bouncing phase. During this regime $z^2 \simeq \frac{a^2 3\beta }{\gamma^2\dot{\phi}^2}$ and $c_s^2 \simeq-1/3$ from \eqref{eq:bz},\eqref{eq:bcs}. A negative sound speed square leads to an exponential growth of the perturbation modes. However, the bounce is short enough so that the evolution of modes remains under perturbative control. All modes are amplified by the same finite amount. This is further verified by the numerical study.
For super-horizon modes
an approximate solution to the perturbation equations near the bounce is given by
\begin{equation}
    v_k(\tau) \simeq d_{1}(k) e^{\omega(\tau-\tau_{B-})} +  d_{2}(k) e^{-\omega(\tau-\tau_{B-})}
\end{equation} 
where 
\begin{equation}
\begin{split}
    tensor: \quad \omega^2 &\simeq \Upsilon    \\
   scalar: \quad \omega^2 &\simeq \Upsilon +\frac{2}{T^2} +\left(2\Upsilon^2+\frac{6\Upsilon}{T^2}+\frac{4}{T^2} \right) t^2 
    \end{split}
\end{equation}
where $T$ is approximately one-quarter of the duration of the bounce \cite{Cai:2012va}. Note that $\Upsilon \ll T^{-2}$ which is why
 tensor perturbations remain unchanged across the bounce while scalar fluctuations are amplified by a factor of $e^{\int_{B-}^{B+} \omega d\tau}$. For scalar perturbations, this factor is \cite{Cai:2007zv,Cai:2008ed,Cai:2008qw}
\begin{equation}
    F_{\zeta} = e^{\int_{\tau_{B-}}^{\tau_{B+}} \omega d\tau} \simeq e^{\left( \sqrt{\Upsilon+\frac{2}{T^2}} t+\frac{2+3 \Upsilon T^2+ \Upsilon^2 T^4}{3T^4\sqrt{\Upsilon+\frac{2}{T^2}}} t^3 \right)|_{\tau{B-}}^{\tau_{B+}}}.
\end{equation}
 For tensor perturbations, amplification is small since $\omega$ is a small constant, $F_h\simeq 1$. 
This growth rate is approximately the same for all infrared modes and is $k-$independent.
\subsection{Kinetic domination}
Since $\phi$ has a large positive velocity at the bounce point, it continues to increase after the bounce. Within a short time, it will cross the second boundary of the ghost
condensation region $\phi_{B^+}$. At that point, the Lagrangian
of the model recovers the canonical form and the universe
enters a kinetic-driven phase of expansion. After the end of the kinetic-driven expansion, the standard radiation-dominated, matter-dominated, and Dark Energy dominated epochs ensue. %The end of the kinetic expansion and beginning of radiation dominated era is denoted by conformal time  $\tau_r$
During kinetic domination, the mode functions are
\begin{equation}
     v_k(\tau)\simeq e_1(k) H^{(1)}_0(k (\tau-\tau_{B+})) +  e_2(k) H^{(2)}_0(k (\tau-\tau_{B+})).
\end{equation}
\subsection{Matching conditions and solutions}
Having obtained solutions to perturbation equations at all three phases, one can match the solutions at appropriate transition surfaces. Tensor perturbations are well-behaved across the bounce as the speed of gravitational waves remains unchanged across the bounce. Hence, it is enough to verify that $h_k$ and $h_k'$ are continuous at the transition points.\\
In general scalar fluctuations are more problematic. Instabilities can arise due to $c_s^2$ being negative.
It has been shown in \cite{PhysRevD.28.679,Brandenberger:1983tg} that in the absence of entropy fluctuations the curvature perturbation $\zeta$ in constant field gauge is a conserved quantity on large scales in an expanding Universe. However, in the context of bounce,  $\zeta$ is not necessarily well-behaved. Nevertheless, in our model it has been shown that the uniform field gauge is well-behaved throughout the evolution of the Universe, and matching conditions for the background are satisfied on both transition surfaces \cite{Cai:2012va}. Hence, matching conditions in \cite{Hwang:1991an,Deruelle:1995kd} apply, indicating that $v$ and $v'$ are continuous across the surfaces, i.e. at $\tau_{B-}$ and $\tau_{B+}$. The growing mode of perturbations dominates at later times.
The primordial spectrum at the end of the kinetic expansion epoch is given by \cite{Cai:2012va},
%\begin{equation}
   % P_{\zeta,h} = F_{\zeta,h}^2 \left(\frac{k}{H_{B^-}} \right)^{\frac{6(1+w)}{1+3w}} %\frac{2^{\frac{4}{1+3w}}\gamma_{E}^2 H_{B^-}^2}{48 \pi^2 (1+w)^{\frac{4}{1+w}}M_p^2}.
%\end{equation}
\begin{equation}
    P_{\zeta} = \frac{k^3}{2\pi^2} \left( \frac{\abs{v_{k}}}{z} \right)^2, \;\;\;\;\;\;\;\;\;\; P_{h} = \frac{4k^3}{\pi^2} \left(\frac{\abs{v_{k}}}{a} \right)^2.
\end{equation}
Considering the vacuum initial condition
\begin{equation}
    v_k(\tau) \longrightarrow \frac{e^{-i k (\tau-\tau_{B-})}}{\sqrt{2k}}
\end{equation}
and implementing the matching conditions, $v_k$ at the end of the kinetic expansion epoch is given by \cite{Cai:2012va}
\begin{equation}
   v_{k} = F_{\zeta,h}  \frac{\gamma_{E} a(\tau)}{2^{\frac{3}{2}}k^{\nu}(2H_{B-})^{\nu-\frac{1}{2}}}
\end{equation}
where $\nu = \frac{3}{2} \frac{w-1}{3w+1}$.
Hence the tensor and scalar perturbations are given by 
\begin{equation}
    P_{h} = \left(\frac{k}{H_{B^-}} \right)^{\frac{6(1+w_c)}{1+3w_c}} \frac{\gamma_{E}^2 H_{B^-}^2}{2 \pi^2 M_p^2}. 
\end{equation}
($F_{h}$=1) and
\begin{equation}
    P_{\zeta} = \left(\frac{k}{H_{B^-}} \right)^{\frac{6(1+w_c)}{1+3w_c}}  \frac{\gamma_{E}^2 H_{B^-}^2}{48 \pi^2 M_p^2} F_{\zeta}^2. 
    \label{eq:nvacs}
\end{equation}
 Where $\gamma_{E}$ is the Euler-Mascheroni constant and $ F_{\zeta}^2$ is the amplification factor for the scalar power spectrum. $w_c$ is the equation of state during ekpyrotic contraction and is defined in the earlier sections. $w_c\gg1$ thus, we find that $n_s-1=2$ and $n_T=2$ indicate a blue tilted spectrum as expected from ekpyrotic scenarios.  It is interesting to notice that the amplification of the scalar power spectrum could lead to interesting predictions regarding the vacuum spectrum for future gravitational wave observations such as LISA and LIGO. We leave this discussion for another time and focus on the sourced perturbations and their evolution across the bounce.
\section{Sourced tensor perturbations} 
\label{sec:sourced}
%We have established that the background evolution of the universe is non-singular and stable and calculated the vacuum perturbations across the bounce. As expected, they match behave as expected in the absence of gauge fields, 
After deriving the behaviour of the background and vacuum fluctuations, we are interested in studying perturbations sourced by gauge fields and their evolution across the bounce. This evolution is of major importance as in our model, the sourced fluctuations are the nearly scale-invariant ones measured on CMB scales. In previous works \cite{Artymowski:2020pci,r1,r3} evolution of perturbations stopped at the end of the ekpyrotic/contracting phase under the assumption that the bounce does not cause significant changes in the spectra. %and hence equations for gauge fields remained the same throughout the evolution. %We have assumed that the short bounce does not cause any significant changes in the power spectrum of tensor and scalar perturbations. If this were true, perturbations from the contracting regime seed the inhomogeneities in the CMB and their power spectrum can be compared to observations. However,
We have just demonstrated that for a non-singular bounce, the amplitude of the scalar power spectrum is amplified by a significant factor as the perturbations evolve across the bounce. We expect similar amplification of the spectrum from sourced perturbations. %This amplification of the scalar spectrum can drive the tensor-to-scalar ratio to be small enough such that perturbations sourced by U(1) gauge field are inside the viable regime of models. 
For that purpose, we need to solve for the gauge fields, the tensor modes, and the scalar modes across the bounce. %When approaching the bounce, the coupling between the scalar changes significantly from the usual $\tau^n$ coupling, attaining a constant value, equation for the gauge field is different near the bounce. 
Nonetheless, we will show that the scale invariance of the source spectrum is unaffected by the bounce.

%\subsection{Tensor perturbations} \label{subsec:tensor}
The tensor source term $J_{\lambda}^T (\tau,\Vec{k})$ is obtained by taking the transverse and traceless spatial part of the energy-momentum tensor and projecting it along the $\lambda$ polarization. For our model of gauge fields \cite{r3}
\begin{equation}
\begin{split}
J_{\lambda}^T (\tau,\Vec{k}) &\simeq -\frac{1}{2 a}\int \frac{d^3p}{2\pi^3} \sum_{\lambda'} \epsilon_i^{\lambda*}(\vec{k}) \epsilon_i^{\lambda*}(\vec{k})
\epsilon_i^{\lambda'}(\vec{p})\epsilon_j^{\lambda'}(\vec{p}-\vec{k}) \frac{\left(\tilde{A}'_{\lambda} I -\tilde{A}_{\lambda} I'\right)}{I^2}\\
&\;\;\;\;\;\;\;\; \times \left[ \hat{a}_{\lambda'}(\vec{p})+\hat{a}^{\dagger}_{\lambda'}(-\vec{p})\right]\left[ \hat{a}_{\lambda'}(\vec{k}-\vec{p})+\hat{a}^{\dagger}_{\lambda'}(-\vec{k}+\vec{p})\right].
\end{split}
\label{eq:tensource}
\end{equation}
Equations for tensor perturbations are unaffected by the intricacies of the background field. Hence, Fourier modes are given by the equation 
\begin{equation}
      h''(k,\tau)+\left( k^2-\frac{a''}{a} \right) h(k,\tau) = J_{\lambda} (\tau,\Vec{k}) \, .
    \label{eq:muksasts}
\end{equation}
As we have seen in previous works, one of the polarizations will have a significantly larger amplitude \cite{r3}.
In order to evolve perturbations through the bounce, we note that $I(\tau)$ is approximately a constant throughout the duration of the bounce, equation for the gauge field \eqref{eq:guage} in this regime can be approximated as 
\begin{equation}
\tilde{A''}_{\lambda}+k^2 \tilde{A}_{\lambda} = 0 \,  \label{eq:guageb}.
\end{equation}
Gauge fields behave as though they are in the Minkowski vacuum, and their solution is accordingly
%We assume that solutions are Minkowski during the bouncing phase. ie
\begin{equation}
     \Tilde{A_{\lambda}}(k,\tau) = c1(k) e^{ik\tau} + c2(k) e^{-ik\tau}.
\end{equation}
This solution needs to be matched with the solution in the contracting phase, where outside the horizon ($k\tau \ll 1$) is approximated by \cite{r3} 
\begin{equation}
\tilde{A}(k,\tau) = -\sqrt{\frac{-\tau}{2\pi}} e^{\pi \xi} \Gamma(\abs{-2n+1})\abs{ 2 \xi k \tau}^{-\abs{-n+\frac{1}{2}}} \ .\label{gaugefield}
\end{equation} 
We restrict the equation of modes of gauge fields to $n=2,-1$, which will lead to a scale-invariant spectrum for sourced perturbations. We will use $n=2$ for convenience. 
Requiring the continuity of $ \Tilde{A_{\lambda}}(k,\tau)$ and $ \Tilde{A'_{\lambda}}(k,\tau)$ at the point of transition from contracting phase to bouncing phase gives the following solution during the bouncing phase: 
\begin{equation}
%\begin{split}
     \Tilde{A_{\lambda}}(k,\tau)_{bounce} = e^{\pi \xi}  \sqrt{\frac{-\tau_{B-}}{2 \pi }} \frac{\Gamma(3)}{(-2 \xi \tau_{B-})^{\frac{3}{2}}} k^{-\frac{3}{2}} %\\
     %& \;\;\;\;\;\;\;\;\;\;\;\; 
     \left(\cos(k(\tau-\tau_{B-}))- \frac{ \sin(k(\tau-\tau_{B-}))}{(-k \tau_{B-}) } \right) .
     \label{gauge b}
   %  \end{split}
\end{equation}
Note that near and after the bounce $I\simeq1$.
In previous works, we have shown that the terms containing magnetic field  and terms that are higher order in $k\tau$ can be safely ignored while calculating the source term, as well as subhorizon contributions \cite{r1,r3,Artymowski:2020pci}. Upon expanding $A_{\lambda}$ as a function of $k\tau$ and neglecting all such irrelevant terms, we find that the source term during the bouncing phase, calculated from equation \eqref{eq:tensource} is
\begin{equation}
\begin{split}
    J_{\lambda}^T (\tau,\Vec{k})\simeq & \frac{1}{2 a}\int \frac{d^3p}{2\pi^\frac{3}{2}} 
\epsilon_i^{\lambda}(\vec{p})\epsilon_j^{\lambda}(\vec{p}-\vec{k}) \left( e^{2\pi \xi} \frac{ \Gamma(3)^2}{2\pi  (2 \xi p)^{\frac{3}{2}}(2 \xi \abs{\vec{k}-\vec{p}})^{\frac{3}{2}} } \right) \\
&\;\;\;\;\;\;\;\; \left(\frac{1}{-\tau_{B-}} \right)^2 \left[ \hat{a}_{\lambda}(\vec{p})+\hat{a}^{\dagger}_{\lambda}(-\vec{p})\right]\left[ \hat{a}_{\lambda}(\vec{k}-\vec{p})+\hat{a}^{\dagger}_{\lambda}(-\vec{k}+\vec{p})\right].\end{split}
\label{sourceb}
\end{equation}
Initial conditions for the bounce are set by fluctuations at the end of the sourced ekpyrotic contraction. We shall now show analytically that the tensor spectrum is practically unchanged across the bounce.
\subsection{Analytic approximations}
\label{sec:analap}
We follow the approximations employed in \cite{Cai:2012va} to calculate vacuum perturbations in order to derive an analytic approximation for the sourced power spectrum. Near the bounce
$H=\Upsilon t$ and $a \simeq e^{\Upsilon t^2}$ from which we can conclude that near bounce
\begin{equation}
    \frac{a''}{a} \approx \Upsilon
\end{equation}
The vacuum Mukhanov-Sasaki equation is hence given by 
\begin{equation}
    h_k''+(k^2-\Upsilon) h_k = 0. 
\end{equation}
whose super horizon solutions are given by
\begin{equation}
    h_k = c_1 e^{\omega t} + c_2 e^{-\omega t}
\end{equation}
where
\begin{equation}
    \omega \approx \sqrt{\Upsilon}. 
\end{equation}
Sourced perturbations are solutions to the equation
\begin{equation}
h_k'' + \left( k^2-\frac{a''}{a}\right)h_k =  J^T(\tau,\Vec{k} ) \, , \label{eq:vsourcedt}
\end{equation}
 Since Eq. \eqref{eq:vsourcedt} is linear in $h_i$ and creation/annihilation operators of sourced and unsourced fluctuations are uncorrelated \cite{Barnaby:2012xt}, its solutions and their power spectra $\mathcal{P}$ should be linear combinations of vacuum and sourced fluctuations 
\begin{equation}
h_k =  h_{k,v} + h_{k,s}  \qquad \Rightarrow \qquad \mathcal{P}^{tot} = \mathcal{P}^v + \mathcal{P}^s \, 
\end{equation}
where $h_{k,v}$ is the solution to homogeneous equation and  $h_{k,s}$ is solution to the inhomogeneous problem. $h_{k,v}$ corresponds to vacuum perturbations derived in \cite{Cai:2012va}. In order to determine $h_{k,s}$ at the end of the bounce, we divide the time domain into two regimes. Let $\tau_B^{-}$ and $\tau_B^{+}$ be times at which the field enters and exits the ghost condensate state, respectively. This is roughly the time at which the Universe will enter and exit the bouncing phase. Then, the region of time domain $(-\infty,\tau_{B-})$ is the ekpyrotic regime when the Universe undergoes ekpyrotic contraction. Time interval  $(\tau_{B-},\tau_{B+})$ corresponds to the bouncing phase. Let $h_{ekp,s}$ be the solution to the inhomogeneous equation during the ekpyrotic phase
\begin{equation}
h_{ekp,s} = \int_{-\infty}^{\tau_{B-}} d\tau' G_{k}(\tau,\tau') J^T_k (\tau',\Vec{k}) \, ,
\label{eq:vie} 
\end{equation}
where $G_k(\tau,\tau')$ is the Green's function obtained by solving
\begin{equation}
G^{ \prime \prime } + \left(k^2 - \frac{a''}{a}\right)G = \delta(\tau-\tau') \, .
\label{eq:greenfs}
\end{equation}
during the regime of ekpyrotic contraction 
\begin{equation}
\begin{split}
 \frac{a''}{a} &= \frac{b(1+b)}{(\tau-\tau_B^-)^2}  \qquad  -\infty < \tau < \tau_{B}^{-} \\
\end{split}
\end{equation}
where 
\begin{equation}
    b = \frac{q}{(1-q)}. 
\end{equation}
$G_{0k}(\tau,\tau')$ is the familiar Green's functions for sourced perturbations in a contracting Universe given by \cite{Artymowski:2020pci,r4,r1}
  \begin{equation}
\begin{split}
    G^T_{0k}(\tau,\tau') &= i   \frac{\pi}{4} \sqrt{\tau \tau'} %\times \\ & \, 
    \left[H^{(1)}_{\frac{1}{2}+b}(-k\tau) H^{(2)}_{\frac{1}{2}+b}(-k\tau')-H^{(1)}_{\frac{1}{2}+b}(-k\tau')H^{(2)}_{\frac{1}{2}+b}(-k\tau)\right].\cr
    \end{split}
    \label{eq:geensfsten1}
\end{equation}
Let $h_{b,s}$ be the solution to the inhomogeneous equation during the bouncing phase. During this regime, $\frac{a''}{a} = \Upsilon = \omega^2$ and Fourier modes for sourced perturbations are solutions to the equation
\begin{equation}
    h_k'' + \left( k^2-\omega^2 \right) h_k =  J^T(\tau,\Vec{k} ) \, , \label{eq:vsourcedb}
\end{equation}
with initial conditions $h_{b,s}(\tau_{B-}) = h_{ekp,s}(\tau_{B-}) $ and $h'_{b,s}(\tau_{B-}) = h'_{ekp,s}(\tau_{B-})$. \\
We employ the method of Laplace transforms to solve this equation. Fourier modes of sourced perturbations at the end of the bouncing phase are given by (See App.\ref{app:greens})
\begin{equation}
\begin{split}
h_{b,s} &= h_{ekp,s}(\tau_{B-}) \cosh(\omega(\tau-\tau_{B-}))+ \frac{1}{\omega} h_{ekp,s}'(\tau_{B-}) \sinh(\omega(\tau-\tau_{B-})) \\
&\;\;\;\;\;\;\;\;\;\;\;\;\; 
+\frac{\mathcal{K}}{\tau^4_{B-}} \frac{2 \sinh^2\left( \omega \frac{(\tau-\tau_{B-})}{2}\right)}{\omega^2}
\label{eq:vi} 
\end{split}
\end{equation}
From previous calculations \cite{Artymowski:2020pci}, we know that  $h_{ekp,s} \simeq \mathcal{K} \frac{(-\tau)^{2-2n}}{2n-2} = \frac{\mathcal{K}}{2 \tau^2}$ (for $n=2$).  Using this fact
\begin{equation}
%\begin{split}
h_{b,s} (\tau) = h_{ekp,s}(\tau_{B-}) \left[\cosh(\omega(\tau-\tau_{B-}))+ \frac{2}{\omega \tau_{B-}} \sinh(\omega(\tau-\tau_{B-})) 
+\frac{4}{\tau^2_{B-}} \frac{\sinh^2\left( \omega \frac{(\tau-\tau_{B-})}{2}\right)}{\omega^2}\right]
\label{eq:vir} 
%\end{split}
\end{equation}
Near the bounce, for super horizon modes $\omega(\tau-\tau_{B-})\ll 1$, $k(\tau-\tau_{B-})\ll 1$, we obtain at the end of the bounce phase
%\begin{equation}

            % &\;\;\; \;\;\;\;\; \frac{1}{2\omega^2} \left( \cosh(\omega(\tau-\tau_{B-}))-1 \right) \\
        % &+\frac{1}{2(\omega^2+4k^2)} \left( \cosh(\omega(\tau-\tau_{B-}))-\cosh(2k (\tau-\tau_{B-}))  \right)\\
        % &\simeq \frac{-\frac{1}{2} \omega^2 ((\tau-\tau_{B-})^2) }{2\omega^2} + \frac{\frac{1}{2} (\omega^2-4k^2) ((\tau-\tau_{B-})^2) }{2(\omega^2+4k^2)}  \\
        % & = \frac{\omega^2}{(\omega^2+4k^2)} (\tau-\tau_{B-})^2 
    %\end{split}
%\end{equation}
%\begin{equation}
%\begin{split}
%h_{b,s} &= h_{ekp,s}(\tau_{B-}) \cosh(\omega(\tau-\tau_{B-}))+ \frac{2 (\tau-\tau_{B-})}{ \tau_{B-}} h_{ekp,s}(\tau_{B-}) \\
%&\;\;\;\;\;\;\;\;\;\;\;\;\; +h_{ekp,s}(\tau_{B-})\frac{ (\tau-\tau_{B-})^2}{\tau^2_{B-}} 
%\label{eq:vif} 
%\end{split}
%\end{equation} 
%when $\tau=\tau_{B+}$,
%\begin{equation}
 % h_{ekp,s}(\tau_{B-})\frac{ (\tau_{B+}-\tau_{B-})^2}{\tau^2_{B-}} - \frac{2 (\tau_{B+}-\tau_{B-})}{ \tau_{B-}} h_{ekp,s}(\tau_{B-})  \simeq 0
%\end{equation}
\begin{equation}
   h_{b,s} (\tau_{B+})\simeq \left(\frac{\tau_{B+}}{\tau_{B-}}\right)^2 h_{ekp,s}(\tau_{B-})
    \label{eq:greenr}.
\end{equation}
For a symmetric bounce, there is no enhancement, while for an asymmetric bounce, there may be some change, but it will, in general, be small compared to the scalar spectrum that will be enhanced by orders of magnitude.
%and we obtain equation \eqref{eq:greenr}, namely. 
%For future refferenc
%\begin{equation}
 %  h_{b,s} \simeq \cosh(\omega(\tau-\tau_B^-)) h_{ekp,s}(\tau_{B-})
   % \label{eq:greenr}.
%\end{equation}
Sourced perturbations from ekpyrosis are amplified by a factor of $ (\tau_{B+}/\tau_{B-})^2$ as they cross the bounce, this factor is approximately of order one for short bounce and small $\omega$.  
Hence, the sourced tensor spectrum does not change significantly through the bounce. The tensor spectrum after the bounce is the same as the spectrum at the end of the contraction, given by \cite{r3,r4}
\begin{equation}
P^T_S  = \abs{\hat{f^T(q)}}^2  \frac{9 e^{4\pi \xi}}{8 \pi b^4 \xi^6} \left(\frac{H_{B-}}{M_{pl}}\right)^4 \left(\frac{k}{H_{B-}}\right)^{n_T},
\label{eq:ictspectrum}
\end{equation}
for $n\simeq2$, where $H_{B-}$ is the Hubble parameter at $\tau=\tau_{B-}$ and 
\begin{equation}
\abs{\hat{f}^T}^2 = \int \frac{d^3p}{(2\pi)^\frac{3}{2}} \abs{ P_{\lambda}}^2 \left(p\abs{\vec{k}-\vec{p}}\right)^{-3}\\
\end{equation}
where $P_{\lambda}\left(\vec{k},\vec{p},\vec{k}-\vec{p}\right)=\sum_{\lambda} \epsilon_i^{\lambda*}(\vec{k}) \epsilon_j^{\lambda*}(\vec{k})
\epsilon_i^{\lambda'}(\vec{p})\epsilon_j^{\lambda'}$, and  $\abs{\hat{f}^T}$ was calculated in \cite{r4}. \\
 We verify our analysis numerically. The numerical evaluation of the power spectrum agrees with our analysis that the tensor spectrum remains relatively unchanged across the bounce. %Upon examining the numerical simulations, we find that the tensor spectrum remains relatively unchanged across the bounce. 
\begin{figure}[H]
    \centering
    \includegraphics[width=0.45\textwidth]{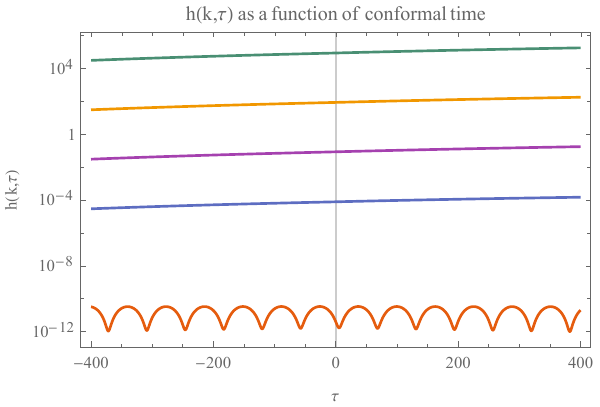} 
    \hspace{0.5cm}
     \includegraphics[width=0.45\textwidth]{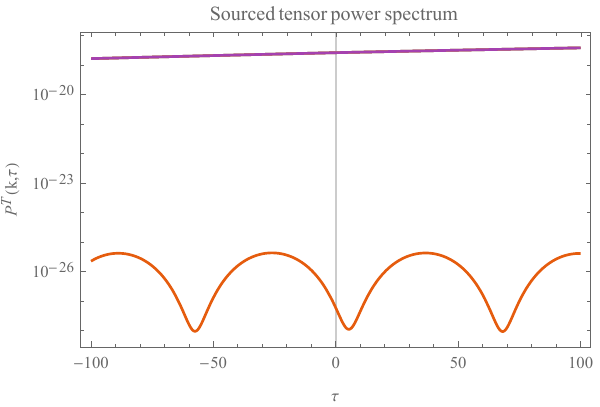} \\
     \includegraphics[width=0.45\textwidth]{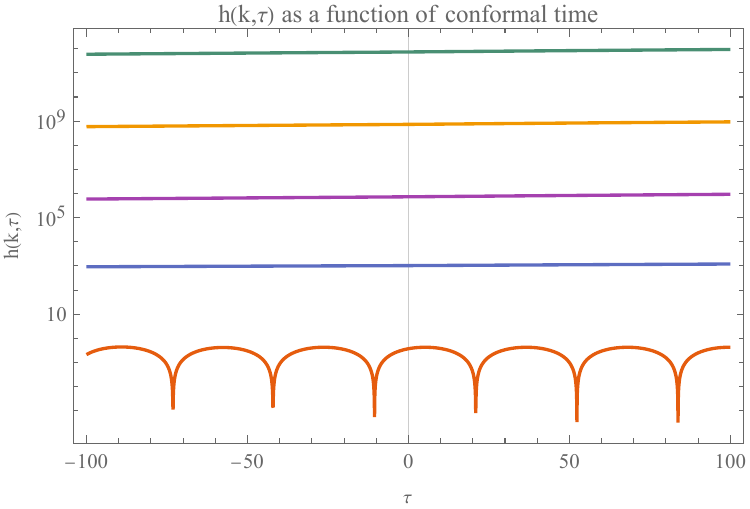} 
    \hspace{0.5cm}
     \includegraphics[width=0.45\textwidth]{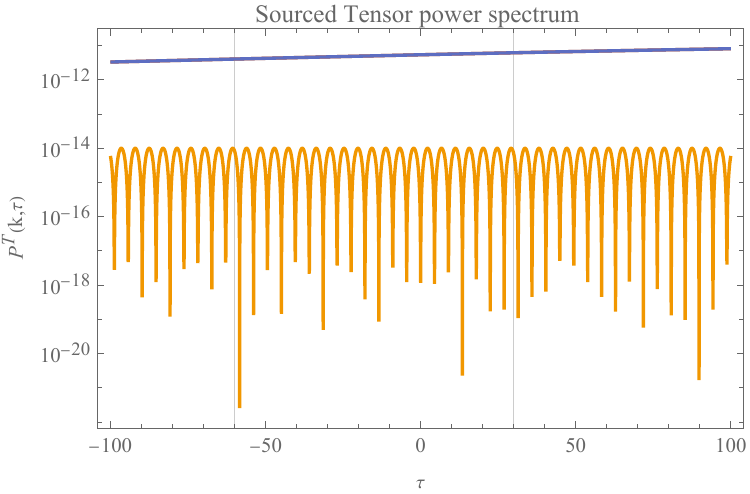} 
     \\
     \caption{Left and right panels show Fourier modes of tensor fluctuations and their power spectrum, respectively for two different sets of parameters (top and bottom, see text). The upper curvess are modes exiting the horizon while the lower one remains subhorizon for all times. The amplitude is changing minimally across the bounce, and the scale dependence of the power spectrum for super-horizon modes is nearly zero.} 
\label{fig:sourced tensor spectrum}
\end{figure}
In figure \ref{fig:sourced tensor spectrum}, we have numerically evaluated the sourced tensor perturbations across the bounce with parameter values 
$V_0 = 10^{-7},g_0 = 1.1,\beta = 5,\gamma = 10^{-3},b_V = 5,
b_g = 0.5, p = 0.05 ,q = 0.1$, energy density of gauge fields at $t=0$ at $10^{-8}$ and $\xi = 2.4$ (Top panels), and $V_0 = 10^{-7},g_0 = 1.1,\beta = 200,\gamma = 10^{-3},b_V = 5,
b_g = 0.3, p = 0.08 ,q = 0.1$, energy density of gauge fields at $t=0$ at $10^{-8}$ and $\xi = 2.4$ (bottom panels). 
The spectrum is practically scale invariant with $n_T = \dfrac{d\log(P_h)}{d\log(k)} \approx 10^{-5}$.
This invariance of the tensor spectrum reassures us that if the scalar spectrum is amplified similarly to vacuum perturbations in a non-singular bounce, this amplification will drive the tensor to scalar ratio of the model to significantly small values. This will solve the  remaining significant drawback of the model of sourced perturbations.

\section{Sourced scalar perturbations}
\label{sec:scalar}
The scalar spectrum of vacuum perturbations in a non-singular bouncing universe is given by \eqref{eq:nvacs}. Spectrum is amplified by  a factor $F_{\zeta}$ compared to the spectrum at the end of the slow contraction. Sourced fluctuations appear at second order in perturbative expansion and are amplified such that they are larger than vacuum perturbations of a lower order, by the presence of an exponential factor in the equation for gauge fields. To find if the sourced perturbations undergo amplification similar to vacuum perturbations, as they cross the bounce we need to derive perturbation equations for second-order scalar fluctuations and solve them. To derive perturbation equations, we need to expand the action in \eqref{eq:vecaction} up to the third order. We derive the third-order action including the source term, and calculate the scalar spectrum across the bounce. We verify our analytical results with numerical ones.
\subsection{Expansion of action}
To derive the full perturbation equations, we write the combined second and third-order action and vary it with respect to $\zeta$, the gauge-invariant curvature perturbations, and expand it afterwards with $\zeta = \zeta_1 + \zeta_2$, where $\zeta_1$ and $\zeta_2$ stand for first and second order perturbations respectively. The second order action for Galileons of our model coupled to gravity  is given by \cite{Gao:2011qe,Cai:2012va}\footnote{In the following pages $\partial$ represent spatial derivatives, $\partial^2 \zeta = \delta^{ij} \partial_i \partial_j \zeta$ and $(\partial_i \zeta)^2 = \delta^{ij} \partial_i\zeta \partial_j \zeta$ etc. }
\begin{equation}
\begin{split}
    S_{2}[\zeta,\alpha,\beta] &= \int d\tau d^3x  \frac{ a^2}{2} \Bigg[-3 \zeta'^2+ (\partial \zeta)^2-3\mathcal{H}^2 m_{\alpha} \alpha^2%\right]
    \\
   \;\;\;\;\;\ &%+ \int d\tau d^3x \frac{a^2}{2}  \left[
  +2 \partial \alpha \partial \zeta + 6\mathcal{H} f_{\alpha} \alpha \zeta'+2 \zeta'\partial^2 \beta -2\mathcal{H} f_{\alpha} \alpha \partial^2 \beta  \Bigg]
    \label{eq:ac2nd}
   \end{split}
\end{equation}
where
\begin{equation}
    f_{\alpha} = 1+ \frac{\phi'^3}{2\mathcal{H}} \gamma
\end{equation}
and
\begin{equation}
    m_{\alpha} = 1-\frac{a^2}{6\mathcal{H}^2} \phi'^2 \left(1-g(\phi)+  3 \beta \phi'^2  \right) -\frac{3 \gamma \phi'^3}{2\mathcal{H}}.
\end{equation}
and $\alpha$ and $\beta$ are ADM constraints. $\alpha$ and $\beta$ are obtained by varying \eqref{eq:ac2nd} with respect to $\alpha$ and $\beta$
\begin{equation}
    \alpha = \frac{1}{\mathcal{H}  f_{\alpha}} \zeta',
    \label{eq:alpha}
\end{equation}
\begin{equation}
\partial^2 \beta = \frac{1}{\mathcal{H}  f_{\alpha}}  \partial^2 \zeta-3\left(\frac{m_{\alpha}}{f_{\alpha}^2}-1 \right) \zeta' .
\label{eq:beta}
\end{equation}
Upon substituting \eqref{eq:alpha} and \eqref{eq:beta} in action \eqref{eq:ac2nd} 
we are left with only terms containing $\zeta'^2$ and $\partial^2 \zeta$. We have used integration by parts to reduce the following  term to the form we desire:
\begin{equation}
    \begin{split}
    & \int  \partial \left(\frac{\zeta'}{f_{\alpha} \mathcal{H}} \right) \partial \zeta = -\frac{1}{2}\int \left(\frac{1}{f_{\alpha} \mathcal{H}} \right)' (\partial \zeta)^2.
    \end{split}
\end{equation}
Collecting all these terms together, we obtain
\begin{equation}
\begin{split}
    S_{2}[\zeta,\alpha,\beta] = \int d\tau d^3x  \frac{1}{2} z^2 \left(\zeta'^2-c_s^2(\partial\zeta)^2 \right)
    \end{split}
\end{equation}
where 
\begin{equation}
    z^2 =  6 a^2 \left(1-\frac{m_{\alpha}}{f_{\alpha}^2} \right) 
\end{equation}
and 
\begin{equation}
    c_s^2 =\frac{1}{3}  \left(\frac{1}{a^2}\left(\frac{a^2}{\mathcal{H} f_{\alpha}}\right)'-1 \right) \left(1-\frac{m_{\alpha}}{f_{\alpha}^2} \right)^{-1} .
\end{equation}
By virtue of expanding $m_{\alpha}$ and $f_{\alpha}$ we have verified that this result is consistent with equations \eqref{eq:bz} and \eqref{eq:bcs}.
\subsection{Third order action}
In order to derive the perturbation equations containing the source term of the second order, we need to vary the second and third-order action. We have already established second-order action and have derived the vacuum perturbation equations in the previous section. 
Third-order action for Galileons was derived in earlier works to study non-Gaussianities in inflation models with Galileons \cite{Gao:2011qe,Koehn:2015vvy}. 
\begin{equation}
\begin{split} 
    S_3[\zeta] &= \int d\tau d^3x  a^2  \left[-9 \zeta \zeta'^2+2\zeta'(\zeta \partial^2(\frac{1}{\mathcal{H}f}\partial^2\zeta-3 \kappa \zeta')+  \partial_i\zeta \partial^i \beta \right]\\
    &-a^2 \left[\frac{1}{\mathcal{H} f} \partial_i \zeta (\partial^i \zeta)^2+ \partial_i((\frac{1}{\mathcal{H} f}\partial^2\zeta-3 \kappa \zeta'))^2 \partial^2 \zeta-\frac{1}{2} \zeta \left(4 \frac{1}{\mathcal{H} f}  \zeta' \partial^2 \zeta \right)-(\partial^2 \beta)^2+ \partial_i \partial_j \beta \right]\\
    & + a^2  \left( \zeta (\partial_i \zeta)^2-9 \mathcal{H}^2 m_{\alpha} \alpha^2 \zeta+ 2\mathcal{H} f_{\alpha} \alpha (9\zeta \zeta'-\zeta \partial^2\beta - \partial_i \zeta \partial^{i} \beta) \right) \\
   & + \frac{\lambda_1}{\mathcal{H}} \left[ \zeta'^3-\zeta'^2\partial^2 \beta +\frac{1}{2} \zeta' \left(4 \alpha \partial^2 \zeta + (\partial^2 \beta)^2-(\partial_i \partial_j \beta)^2-\alpha \left(\partial^2 \zeta \partial^2 \beta -\partial_i \partial_j \zeta \partial_i \partial_j \beta  \right) \right) \right] \\
   & + \lambda_2 \alpha \left[3 \zeta'^2-2\zeta'\partial^2 \beta+ \frac{1}{2} (\partial^2 \beta)^2 - (\partial_i \partial_j \beta)^2 - \lambda_3 \mathcal{H} \alpha^2 (3 \zeta'-\partial^2 \beta)-\lambda_4 \alpha^2 \partial^2 \zeta + \frac{\lambda_5}{2} \mathcal{H}^2 \alpha^3  \right], 
    \end{split}
    \label{eq:ac3rd}
\end{equation}
where $\kappa = 3\left(\frac{m_{\alpha}}{f_{\alpha}^2}-1\right)$. 
Substituting the Galileon limit in equation for $\lambda$s derived in \cite{Gao:2011qe}, we conclude that for our model $\lambda_1=0, \lambda_2=1, \lambda_4=1$, and 
\begin{equation}
    \lambda_3 = 1+ \frac{\phi'}{2\mathcal{H}}  3 \gamma \phi'^2,
    \label{eq:lambda3}
\end{equation}
\begin{equation}
    \lambda_5 = 1+ \frac{a^2}{\mathcal{H}^2} \left((1-g(\phi))\phi'^2 + \beta \frac{\phi'^2}{2}+4 \beta \phi'^4-V(\phi)  \right) +  9 \frac{\phi'^3}{\mathcal{H}} \gamma .
    \label{eq:lambda5}
\end{equation}
Upon taking the variation of $S_2+S_3$ with respect to $\zeta$ and expanding $\zeta$ as $\zeta_1+\zeta_2$ where $\zeta_1$ and $\zeta_2$ are corrections to curvature perturbations of first and second order, we obtain an equation of the form ( See App.\ref{app:second order equations} )
\begin{equation}
    \zeta_2'' z + 2  z' \zeta_2' + z c_s^2 k^2  \zeta_2+ O(\zeta_1^2) \; terms+ Higher\; order\; terms = \frac{J_s(\tau,k)}{z}.
    \label{eq:real3}
\end{equation}
$O(\zeta_1^2)$ terms will only contribute a blue tilted spectrum that is of significantly lower amplitudes than perturbations sourced by gauge fields similar to second-order perturbations sourced by terms containing first-order field perturbations in \cite{r1,r3}. Ignoring such irrelevant second-order terms and the higher-order terms, equations for second-order perturbations are given by
\begin{equation}
     \zeta_2'' z + 2  z' \zeta_2' + c_s^2 k^2  z \zeta_2 = \frac{1}{z}J_s(\tau,k).
     \label{eq:final}
\end{equation}
where $J_s(\tau,k)$ is the source term for scalar perturbations.
\subsection{Sourced scalar spectrum}
We need to determine the form of $J_s$ for our action. 
The sourced part of the ADM action in the uniform field gauge $\delta \phi=0$ and $h_{ij}=a^2 e^{2\zeta} \delta_{ij} $, is given by
\begin{equation}
    \int dt d^3x a^3 e^{3\zeta} (1+\alpha) I^2(\phi) \left( \frac{1}{4}F^{\mu\nu}F_{\mu \nu} -\frac{\delta}{4}\tilde{F}^{\mu \nu}F_{\mu\nu}\right).
\end{equation}
The source term is hence obtained by varying this term with respect to $\zeta$ 
\begin{equation}
    J_s = \fdv{ \int dt d^3x a^3 e^{3\zeta} (1+\alpha) I^2(\phi) \left( \frac{1}{4}F^{\mu\nu}F_{\mu \nu} -\frac{\delta}{4}\tilde{F}^{\mu \nu}F_{\mu\nu}\right) }{\zeta}.
\end{equation}
Expanding this source term further, including the ADM expansion of the metric and excluding the terms containing magnetic fields similar to \cite{Artymowski:2020pci} (Magnetic fields are smaller than electric fields by order of $k\tau$)
\begin{equation}
    \begin{split}
       J_s \simeq \fdv{ \int d\tau d^3x a^4 e^{\zeta} (1-\alpha) I^2(\phi) \frac{\abs{\vec{E}}^2}{2 a^4}}{\zeta}. 
       \label{eq:j5}\\
    \end{split}
\end{equation}
In terms of gauge fields
\begin{equation}
    \abs{\vec{E}^2} = \frac{1}{2 a^4 (2\pi)^\frac{3}{2}}\int d^3p \Sigma_{\lambda} \left( \dfrac{d A_{\lambda}}{d\tau} \right)^2 \hat{\mathcal{P_{\lambda}}}
\end{equation}
where 
\begin{equation}
    \hat{\mathcal{P_{\lambda}}} =  \epsilon_i^{\lambda*}(\vec{k}) \epsilon_i^{\lambda*}(\vec{k})
\epsilon_i^{\lambda'}(\vec{p})\epsilon_j^{\lambda'}(\vec{p}-\vec{k}) \left[ \hat{a}_{\lambda'}(\vec{p})+\hat{a}^{\dagger}_{\lambda'}(-\vec{p})\right]\left[ \hat{a}_{\lambda'}(\vec{k}-\vec{p})+\hat{a}^{\dagger}_{\lambda'}(-\vec{k}+\vec{p})\right].
\end{equation}
where $A_{\lambda}=\frac{\tilde{A}_{\lambda}}{I}$.
Thus equation \eqref{eq:j5} can be rewritten as 
\begin{equation}
    \begin{split}
       J_s = \fdv{\frac{1}{2 (2\pi)^\frac{3}{2}}\int d^3k \int d\tau d^3x  \Sigma_{\lambda}e^{\zeta} (1-\alpha) I^2(\phi) \left( \dfrac{d A_{\lambda}}{d\tau}\hat{\mathcal{P_{\lambda}}} \right)^2}{\zeta}.  \\
    \end{split}
\end{equation}
Upon performing variation with respect to $\zeta$
\begin{equation}
\begin{split}
    J_s &= \frac{1}{2 (2\pi)^\frac{3}{2}}\int d^3p \Sigma_{\lambda}\left( \dfrac{d}{d\tau} \left(-\frac{I^2}{f_{\alpha}\mathcal{H}} \left( \dfrac{d A_{\lambda}}{d\tau} \right)^2 \right) -  I^2 \left( \dfrac{d A_{\lambda}}{d\tau} \right)^2\right)\hat{\mathcal{P_{\lambda}}}\\
    \end{split}
    \label{eq:sourcece}
\end{equation}
which should be substituted into \eqref{eq:final}.
In the regime of ekpyrotic contraction (See App. (\ref{app:C})),
%\begin{equation}
%\begin{split}
 %    J_s &\simeq  \frac{1}{2 (2\pi)^\frac{3}{2}}\int d^3k \left(\epsilon\left( I^2- \dfrac{d I^2}{d\tau} \frac{1}{\mathcal{\epsilon H}}\right)  \left(\dfrac{d A_{\lambda}}{d\tau}\right)^2 \right) \hat{\mathcal{P_{\lambda}}} \\
  %   &= \frac{1}{2 (2\pi)^\frac{3}{2}}\int d^3k 2\epsilon C(\tau) I^2 \left(\dfrac{d A_{\lambda}}{d\tau}\right)^2 \hat{\mathcal{P_{\lambda}}},
   %  \end{split}
%\end{equation}
%where we labeled the term $\left( \frac{I^2}{2}- \dfrac{d I^2}{d\tau} \frac{1}{2\mathcal{\epsilon H}}\right)$ as $C(\tau) I^2$. 
%We know that sourced term in the equation for perturbations is  $J_{\lambda}^S (\tau,\Vec{k}) = \frac{J_s}{z(\tau)}$, expanding this term and expressing in terms of $\tilde{A}$ we obtain
\begin{equation}
 \begin{split}
J_{\lambda}^S (\tau,\Vec{k})\simeq & \frac{1}{2 a}\int \frac{d^3p}{(2\pi)^\frac{3}{2}} 
\epsilon_i^{\lambda}(\vec{p})\epsilon_i^{\lambda}(\vec{p}-\vec{k}) I^2 \left(\partial_{\tau}\left(\frac{\tilde{A}}{I}\right)\right)^2 \\
&\;\;\;\;\;\;\;\;  \sqrt{2\epsilon(\tau)} C(\tau) \left[ \hat{a}_{\lambda}(\vec{p})+\hat{a}^{\dagger}_{\lambda}(-\vec{p})\right]\left[ \hat{a}_{\lambda}(\vec{k}-\vec{p})+\hat{a}^{\dagger}_{\lambda}(-\vec{k}+\vec{p})\right],
\end{split}
\end{equation}
and\footnote{This $C(\tau)$ during the ekpyrotic regime is consistent with $C(\tau)$ obtained in previous works \cite{Artymowski:2020pci}, where we corrected the constant coefficient to be $1/2$ and not $1$.}
\begin{equation}
C(\tau)= \frac{\mathcal{H}}{\phi'}\left[\dfrac{dI^2}{d\phi}+ \frac{\phi'}{2\mathcal{H}} I^2 \right] I^{-2}(\phi) =  \frac{1}{2}-\frac{I'}{I} \frac{1}{\epsilon \mathcal{H}} = \frac{1}{2} - \frac{d\log I}{d \log H}. 
\label{eq:C1}
\end{equation}
The analysis of the perturbations in the ekpyrotic phase has been carried out in \cite{r1,r3,Artymowski:2020pci}, so we continue with the analysis of the bounce phase. 
As we have shown, gauge field solutions are Minkowski during the bouncing phase i.e.
\begin{equation} \label{eq:Aminkowski}
     \Tilde{A_{\lambda}}(k,\tau) = c1(k) e^{ik\tau} + c2(k) e^{-ik\tau}.
\end{equation}
During the bouncing phase, $f_{\alpha}\neq1$ however, 
 $I$ is a constant. Defining
\begin{equation}
    C_{bounce}(\tau) = \frac{- \left( \dfrac{d}{d\tau} \left(\frac{1}{{f_{\alpha}\mathcal{H}}} \right)\right) }{z(\tau)}.
    \label{eq:c3}
\end{equation}
and substituting \eqref{eq:Aminkowski} results in the following source term (with $n\simeq 2$), %Analysis similar to section \ref{sec:analap} shows that 
\begin{equation}
\begin{split}
    J_{\lambda}^S (\tau,\Vec{k})\simeq & \frac{1}{2 a}\int \frac{d^3p}{(2\pi)^\frac{3}{2}} 
\epsilon_i^{\lambda}(\vec{p})\epsilon_j^{\lambda}(\vec{p}-\vec{k}) C_{bounce}(\tau)  \left( e^{2\pi \xi} \frac{ \Gamma(3)^2}{2\pi  (2 \xi p)^{\frac{3}{2}}(2 \xi \abs{\vec{k}-\vec{p}})^{\frac{3}{2}} } \right) \\
&\;\;\;\;\;\;\;\; \left(\frac{1}{-\tau_{B-}} \right)^2 \left[ \hat{a}_{\lambda}(\vec{p})+\hat{a}^{\dagger}_{\lambda}(-\vec{p})\right]\left[ \hat{a}_{\lambda}(\vec{k}-\vec{p})+\hat{a}^{\dagger}_{\lambda}(-\vec{k}+\vec{p})\right] .
\end{split}
\label{eq:sourceb}
\end{equation}
Now upon substituting $v=z\zeta$ in equation \eqref{eq:final} we obtain
\begin{equation}
    v''+(c_s^2 k^2-\frac{z''}{z}) v =  J_{\lambda}^S (\tau,\Vec{k}).
    \label{eq:vsources}
\end{equation}
We solve equation \eqref{eq:vsources} employing the same method of Laplace transforms used in section \ref{sec:analap}. We divide the time required for the evolution of the Universe into two intervals,$(-\infty,\tau_{B-})$ and $(\tau_{B-},\tau_{B+})$. The vacuum solutions $v_{k,v}$ and sourced perturbations $v_{k,s}$ are independent and hence uncorrelated. 
In the regime of ekpyrosis $(-\infty,\tau_B^-)$ solution to the inhomogeneous equation $v_{ekp,s}$ 
is given by
\begin{equation}
v_{ekp,s} = \int_{-\infty}^{\tau_{B-}} d\tau' G^S_{k}(\tau,\tau') J^S_k (\tau',\Vec{k}) \, ,
\label{eq:vis} 
\end{equation}
where $G^S_k(\tau,\tau')$ is the Green's function obtained by solving
\begin{equation}
G^{S \prime \prime } + \left(c_s^2 k^2 - \frac{z''}{z}\right)G^S = \delta(\tau-\tau') \, .
\label{eq:greenfsg}
\end{equation}
during the regime of ekpyrosis. Green's functions for ekpyrosis and $v_{ekp,s}$ are evaluated in \cite{Artymowski:2020pci,r4}. In order to evolve the perturbations across the bounce, we use WKB approximations. Let $v_{b,s}$ be solution to perturbation equations during the bounce regime $(\tau_{B-},\tau_{B+})$. During the bounce regime for superhorizon modes $c_s^2 k^2 \ll \frac{z''}{z}$ \cite{Cai:2012va}. Within this approximation equation for Fourier modes for curvature perturbations during the bounce regime is given by
\begin{equation}
     v''-\omega_S^2 v =  J_{\lambda}^S (\tau,\Vec{k}) 
\end{equation}
where $\omega_S= \frac{z''}{z}$. Unlike tensor perturbations where $\omega$ was approximately constant, $\omega_S$ varies as a function of time. 
%Hence, the WKB approximation solution for Fourier modes after the bounce $v_{b,s}$ are 
Using the WKB approximation, we obtained $v_{b,s}$ by replacing $\omega_S(\tau-\tau_{B-})$ in equation \eqref{eq:vi2} with $\int_{\tau_{B-}}^{\tau} \omega_S d\tau$. 
\begin{equation}
\begin{split}
v_{b,s} &= v_{ekp,s}(\tau_{B-}) \cosh\left(\int_{\tau_{B-}}^{\tau} \omega_S d\tau \right)+ \frac{1}{\omega_S} v_{ekp,s}'(\tau_{B-} ) \sinh\left(\int_{\tau_{B-}}^{\tau} \omega_S d\tau\right) \\
&+\frac{\mathcal{K}}{\tau^2_{B-}} \frac{\sinh\left(\int_{\tau_{B-}}^{\tau} \omega_S d\tau\right)-\int_{\tau_{B-}}^{\tau} \omega_S d\tau}{\omega_S^3}.\\
\label{eq:viss} 
\end{split}
\end{equation}
Near the bounce, for superhorizon modes $k(\tau-\tau_{B-})\ll 1$, all terms except the first term in \eqref{eq:viss} are insignificant.  It is easier to extract the dominant term here since, for $\omega>1$, the $\cosh(\omega_S)$ term is much larger than the rest. Thus, sourced scalar modes at the end of the bounce are given by
\begin{equation}
   v_{b,s} = \cosh\left(\int_{\tau_{B-}}^{\tau} \omega_S d\tau\right) v_{ekp,s}(\tau_{B-})
    \label{eq:green1}.
\end{equation}
Thus, in general, during a short bounce, amplification across the bounce for sourced perturbations is 
\begin{equation}
F_{\zeta} \simeq \cosh(\int_{\tau_{B}^-}^{\tau_B^+} \omega(\tau'') d\tau'') .
\label{eq:zeta1}
\end{equation} 
We make use of the approximation $z(\tau_{B+})\simeq(\tau_{B-})$ to arrive at a neat expression for curvature perturbations. 
curvature fluctuations $\zeta$ at the end of bounce is $\frac{v_{b,s}}{z(\tau_{B+})} \simeq\frac{v_{b,s}}{z(\tau_{B-})} $, 
\begin{equation}
\zeta(\tau_{B+}) = F_{\zeta}  \frac{v_{ekp,s}(\tau_{B-}) }{z(\tau_{B-})} 
 = F_{\zeta} \; \zeta_{ekp,s}(\tau_{B-}) \, .
\label{eq:timei}
\end{equation}
%Which near the bounce and for super horizon scales can be approximated as 
%\begin{equation}
%\begin{split}
%\mathcal{I}^S & \approx I^S_{v_{k,ekp}} \cosh(\int_{\tau_{B}^-
%}^{\tau{B^+}}  \omega(\tau'') d\tau'') \frac{1}{(1+\omega)^2}  \\
%&+\int d\tau' z(\tau') \frac{1}{(1+\omega)^2}  \tau_B^-  \cosh(\int_{\tau_{B}^-}^{\tau'} \omega(\tau'') %\end{split}
%label{eq:timei}
%\end{equation}
%$\zeta = \frac{v}{z}$ hence
\subsection{Amplification of spectrum across bounce}
\label{subsec:amps}
In order to analytically evaluate the change in the scalar power spectrum across the bounce, we require an approximate analytic expression for $z$ near the bounce. In \cite{Cai:2012va}, this is achieved by approximating the scalar field by a Gaussian near the bounce. However, this does not adequately explain our numerical results. To better explain the behavior of the spectrum across the bounce, we try to solve the background equations of motion \eqref{eq:eom}. 
Near the bounce the first two terms in equation \eqref{eq:eom} are dominant over $V,_{\phi}$ and 
\begin{equation}
   \frac{\ddot{\phi}}{\dot{\phi}}\simeq - \frac{\mathcal{D}}{\mathcal{P}} \simeq \frac{\frac{1}{2} g,_{\phi} \dot{\phi}}{  {1-g + 3 \beta \dot{\phi}^2  }}. 
    \label{eq:dbp}
\end{equation}
We arrive at this approximation from the numerical results we obtain for various parameters we consider.
\begin{figure}[H]
    \centering
    \includegraphics[width=0.45\textwidth]{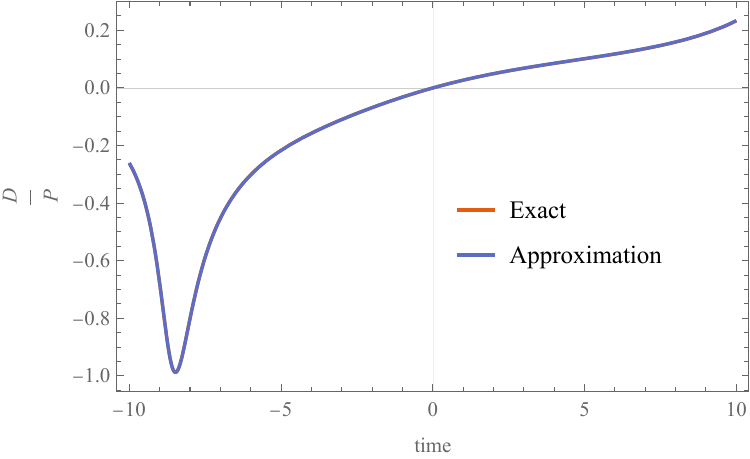} 
    \hspace{0.5cm}
     \includegraphics[width=0.45\textwidth]{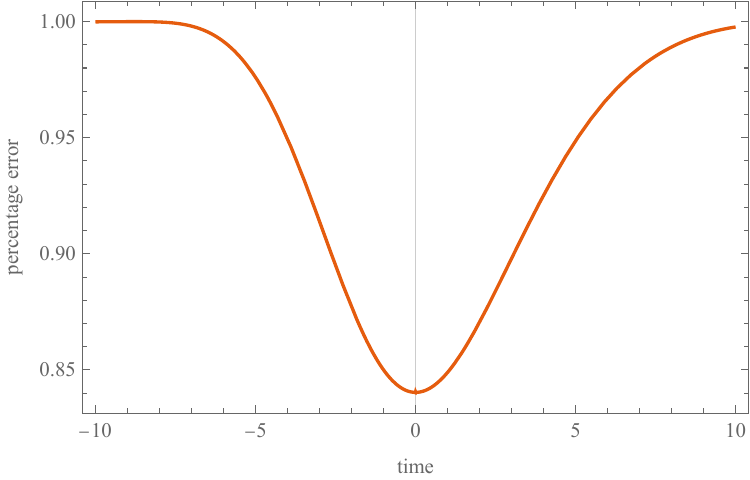} 
     \caption{Left panel show $\frac{\mathcal{D}}{\mathcal{P}}$ and the approximate expression in equation \eqref{eq:dbp} near the bounce for one set of parameters while the right panel depicts the percentage error in the approximation.  We have verified that the percentage error is significantly small for other parameters we have considered. } 
 \label{fig:dp approx}
\end{figure}    

\begin{figure}[H]
    \centering
    \includegraphics[width=0.9\textwidth]{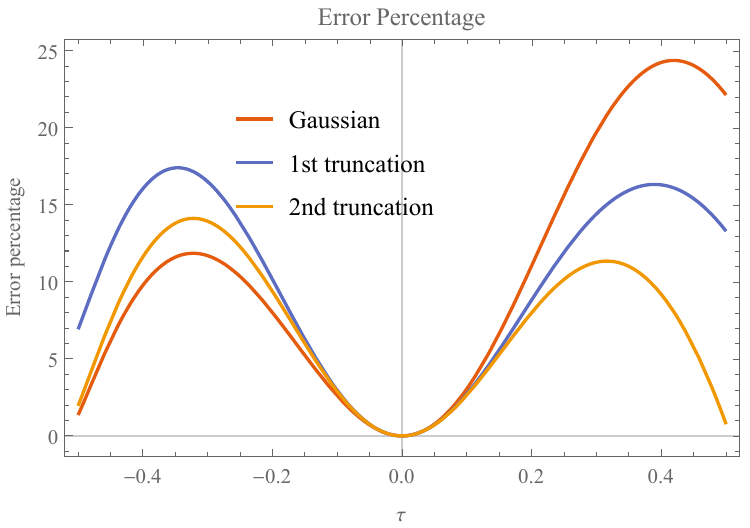} 
     \caption{ Error in approximation for $\phi'(\tau)$ as compared to numerical results as a function of t  near the bounce for the chosen parameters. } 
 \label{fig:comparison}
\end{figure}

%Upon examining various results of numerical calculations, we conclude that the approximation in  \cite{Cai:2012va}
%is insufficient for all the parameters we consider. To illustrate this, with the Gaussian approximation change in amplitude of spectrum across the bounce has a minimum of around $10^8$ however, we show some numerical results for bounce where this is not necessarily the case later on in this thesis. In order to find a new expression for $\dot{\phi}$, 
We start by approximating $\dot{\phi}$ as a Gaussian as the initial solution to the equation of motion, i.e.
\begin{equation}
    \dot{\phi} \approx \dot{\phi}_B e^{-\frac{t^2}{T^2}} 
    \label{eq:skgau}
\end{equation}
where $\dot{\phi_B} = \sqrt{\frac{2(g_0-1)}{3 \beta}}$
and solve the equation perturbatively around the bounce.
%Approximate equations of motion near the bounce are given by 
%\begin{equation}
 %   \ddot{\phi} \approx \dot{\phi} \frac{\mathcal{D}}{\mathcal{P}} \approx \dot{\phi} \frac{\frac{1}{2} g,_{\phi} \dot{\phi}}{  {1-g + 3 \beta \dot{\phi}^2  }}.
   % \label{eq:aeom}
%\end{equation}
%Let the initial solution be $\dot{\phi}_0 =  \dot{\phi}_B e^{\frac{-t^2}{T^2}} $, this solution agrees with numerical calculations within a small range of error for the parameters chosen in \cite{Cai:2012va}. 
We substitute the Gaussian solution back into equation
\eqref{eq:dbp}. Taylor expanding the right-hand side of the equation, we obtain
\begin{equation}
    \frac{\ddot{\phi}}{\dot{\phi}} = -c_1 t + c_2 t^2 + c_3 t^3...
\end{equation}
Exact forms of $c_1,c_2$ and $c_3$ are given in appendix \ref{app:dphi}. 
%We can solve the differential equation \eqref{eq:dbp} perturbatively with $\dot{\phi}_0$ as an initial solution. 
Plugging \eqref{eq:skgau} into \eqref{eq:dbp} at first order in $t$, we identify $c_1=\frac{2}{T^2}$. % should be identified as $\frac{2}{T^2}$ to maintain consistency. 
%Since for the parameters selected in \cite{Cai:2012va} higher order terms are small near the bounce, we are left with only the linear term. For these parameters $c_1=7.9$ and $c_2=3.25$, thus we get back the result $T\approx0.5$ as stated in the paper. 
We compare our approximations to numerical results in figure $\ref{fig:comparison}$.
%In  order for the higher-order term to be small
%\begin{equation}
    %\frac{c_3}{c_2} T < 1 
%\end{equation}
%Largest term in the expansion for $\frac{c_3}{c_2} T$ is
%\begin{equation}
   % \sqrt{\frac{2}{3}} \frac{\sqrt{-1+4g_0}}{3 g_0\sqrt{b_g}}
%\end{equation}
%Hence the expansion remain valid approximately as long as $b_g>\frac{8}{27}$.
Following the identification of $c_1$ we solve equation \eqref{eq:dbp} including the higher order terms iteratively. As a result, we modify  
\eqref{eq:skgau} as
\begin{equation}
    \dot{\phi}(t) = \dot \phi_B e^{-\frac{t^2}{T^2}+ \sigma_1 t^3+ \sigma_2 t^4}
    \label{eq:newphi}
\end{equation}
where $\sigma_1 = \frac{c_2}{3}$ and $\sigma_2=\frac{c_3}{4}$. 
Near the bounce, $z$ can be approximated as 
\begin{equation}
    z^2 \simeq \frac{a^2 3\beta }{\gamma^2\dot{\phi}^2}. 
\end{equation}
From the last two expressions and assuming $\Upsilon$ is small, 
\begin{equation}
%\begin{split}
    \omega_S^2   \simeq \frac{z''}{z} %\\
      \simeq \frac{2}{T^2}- 6 t \sigma_1-12\frac{\sigma_1}{T^2} t^3+24 \sigma_1 \sigma_2 t^5+\left(4\frac{1}{T^4}+9 \sigma_1^2- \frac{16\sigma_2}{T^4} \right)t^4-12\sigma_2 t^2+\sigma_2^2 t^6,
%\end{split}
\label{eq:omegas}
\end{equation}
and amplification across the bounce within the WKB approximation is \eqref{eq:zeta1},
\begin{equation}
    F_{\zeta}= \cosh(\int_{\tau_{B}^-}^{\tau_{B}^+} \omega_S(\tau'') d\tau'') 
     \label{eq:amp}
\end{equation}
The tensor-to-scalar ratio $r$ for the bounce is then given by
\begin{equation}
    r = \frac{P_T^S}{P_S^S} = \frac{P_T^S}{P_S^S}|_{ekpy} \left( \frac{F_{h}}{F_{\zeta}} \right)^2 = \frac{4}{25}  \left( \frac{1}{F_{\zeta}} \right)^2.
\end{equation}
Here we have used our previous result that the tensor-to-scalar ratio during ekpyrosis is $4/25$ \cite{Artymowski:2020pci,r3}. 
Upon substituting the parameters from \cite{Cai:2012va}, we obtain amplification of order $10^{10}$ ($F_{\zeta}\sim 10^5, \, r\sim 10^{-10}$), as expected from the results of our numerical analysis for the same parameters, see figure \ref{fig:sourced scalar spectrum}. %Note that the duration of the bounce is different given this form of $\phi$. The beginning and end of the bouncing phase are  obtained as real solutions to the equation  
%\begin{equation}
    %\frac{t^2}{T^2}- \sigma_1 \frac{t^3}{T^3} + \sigma_2 \frac{t^4}{T^4}= 4. 
   % \label{eq:tb}
%\end{equation}
%\IB{But do you want $\dot{\phi}/\dot{\phi_B}=e^{-4}$ or $\phi/\phi_B=e^{-4}$?}
 % We can obtain four solutions to equation \eqref{eq:tb} assuming $\sigma_2= f \sigma_1$ where $f$ is a real constant. Solutions to equation \eqref{eq:tb} are given in appendix \ref{beb}. 
% One can perform the integral \eqref{eq:amp} to obtain the amplitude increase of the spectrum during the bounce. Tensor to scalar ratio for sourced perturbations is given by
Let us now systematically analyze the different possibilities of amplification and r in the parameter space. %for the amplification of the scalar spectrum in the following sections.
\subsubsection{Symmetric bounce}
We first analyse the case where the bounce is symmetric. For a completely symmetric bounce $b_g=1$. In this case 
\begin{equation}
    \begin{split}
        T = \sqrt{\frac{p \beta}{g_0}}
    \end{split}
\end{equation}
and $\sigma_1=0$, $\sigma_2\simeq0$.
The beginning and end of the bouncing phase are determined by the fact that $g(\phi)$ is larger than one during the bounce phase. At the beginning of the bounce $g(\phi_{B-})=1$ and $\phi_{B-} \simeq \sqrt{\frac{p}{2}}\log(2g_0)$, similarly   $\phi_{B+} \simeq \sqrt{\frac{p}{2}} \frac{\log(2g_0)}{b_g}$. When $b_g=1$ , $\phi_{B+}=\phi_{B-} \simeq \sqrt{\frac{p}{2}}\log(2g_0)$. The field $\phi$ during the bouncing phase is
\begin{equation}
    \phi = \int \dot{\phi}_B e^{-\frac{t^2}{T^2} } dt = \sqrt{\frac{p \pi (-1+g_0)}{6 g_0}} erf\left(\sqrt{\frac{g_0 }{\beta p}} \right)
    \label{eq:phisym}
\end{equation}
where $erf$ is the error function. From equation \eqref{eq:phisym} we conclude that during the bounce phase $\phi < \sqrt{\frac{p \pi (-1+g_0)}{6 g_0}}$. For consistency $\phi_{B+} <\sqrt{\frac{p \pi (-1+g_0)}{6 g_0}}$. This equation does not have a solution for any $g_0>1$ in the real line. Thus, in a completely symmetric bounce, our analytic approximations are insufficient. %not possible to obtain within our framework. 
\subsubsection{Asymmetric bounce with small \texorpdfstring{$b_g$}{bg}}
We have already observed that for $b_g=0.5$, it is possible to get a bounce where the scalar spectrum is amplified by a significant amount. However, we examine the case where $b_g$ is small where expressions considerably simplify. %and a if it is possible to get a smaller amplification so that the tensor-to-scalar ratio is within limits observable by the future CMB experiments. 
In the small $b_g$ limit
\begin{equation}
    \begin{split}
        & T = \sqrt{\frac{(1+4g_0)p \beta}{2b_g(-1+g_0)g_0}}  \;\;\;\;\;\;\;\;\;\;\;\;\;\;\;\;\;\;\;\;  \frac{1}{T^2}= \frac{2b_g(-1+g_0)g_0}{(1+4g_0)p \beta} \\
       & \sigma_1 = -\frac{4b_gg_0}{3\sqrt{3}(-1+4g_0)} \left(\frac{-1+g_0}{p\beta}\right)^{\frac{3}{2}} \;\;\;\;\; \sigma_2 = \frac{b_gg_0}{9(-1+4g_0)} \left(\frac{-1+g_0}{p\beta}\right)^{2}
    \end{split}
\end{equation}
We define a parameter $\rho=\left(\frac{-1+g_0}{p\beta}\right)^{\frac{1}{2}}$, then equation \eqref{eq:newphi} can be rewritten as
\begin{equation}
    \dot{\phi} = \dot{\phi}_B e^{-\frac{2b_g g_0}{1+4g_0} \rho^2 \left(t^2 -\frac{2\rho}{3\sqrt{3}} t^3+   \frac{\rho^2}{9} t^4 \right)}
\end{equation}
We demand that the higher-order terms in the expansion remain small and the argument of exponential is convergent. The next higher order term in the brackets inside exponential is given by $\frac{2}{45\sqrt{3}} \rho^3 t^5$. By demanding that this term is small compared to the fourth-order term, we conclude that 
\begin{equation}
    \rho t < \frac{5\sqrt{3}}{2}.
\end{equation}

\begin{figure}[H]
    \centering
    \includegraphics[width=0.45\textwidth]{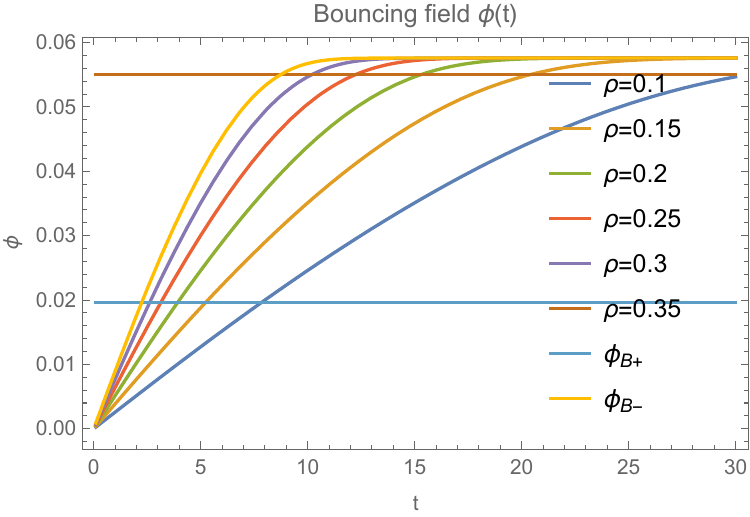} 
    \hspace{0.5cm}
     \includegraphics[width=0.45\textwidth]{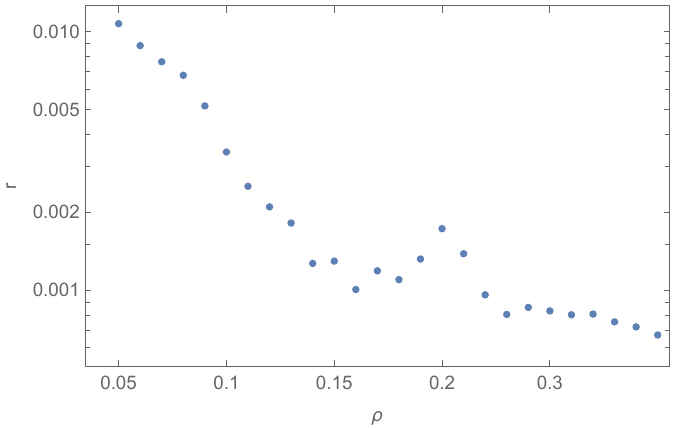} 
     \caption{ Left,  right panels show $\phi$ and $r$ across the bounce for different values of $\rho$, when $p=0.1$, $g_0=1.2$ and $b_g=0.3$.} 
 \label{fig:r}
\end{figure}

\begin{figure}[H]
    \centering
    \includegraphics[width=0.9\textwidth]{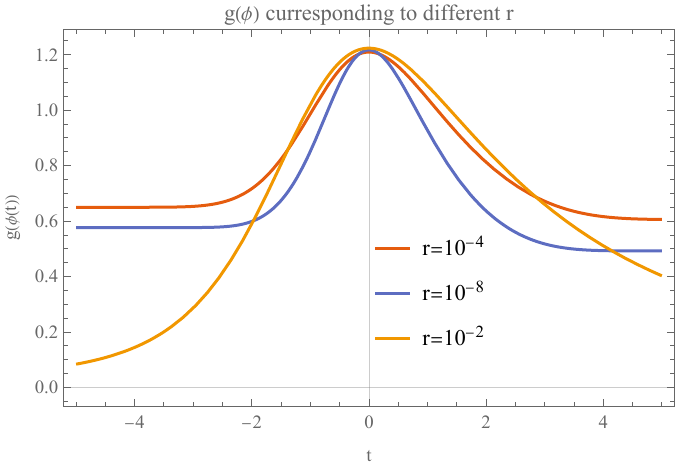} 
     \caption{A depiction of $g(\phi)$ for parameters corresponding to different values of r} 
 \label{fig:g}
\end{figure}

%where $F_{\zeta}\simeq \cosh(\int_{\tau_{B}^-}^{\tau_{B}^+} \omega(\tau'') d\tau'')$.
 %Here we have used our previous result that $r$ for sourced perturbations in the ekpyrotic Universe is $\frac{4}{25}$ \cite{r4,Artymowski:2020pci}.  
 We determine the beginning and end of bounce by plotting the function $\phi=\int \dot{\phi}dt$ with time and deduce the points where $\phi$ is matching $\phi_{B+}$ and $\phi_{B-}$. All integrals are performed numerically. We show the results of our calculations in the left panel of the figure \ref{fig:r}.
In the right panel of figure \ref{fig:r} we show $r$ as a function of $\rho$ (actually $\beta$) with $p=0.1$, $g_0=1.2$ and $b_g=0.3$. Clearly, it is possible to obtain viable and observable r. Depending on the parameter values of choice, it is possible to obtain r ranging from observable to very small. %We show some examples below. 
%\begin{figure}[H]
   % \centering
   % \includegraphics[width=0.45\textwidth]{phit1.pdf} 
  %  \hspace{0.5cm}
     %\includegraphics[width=0.45\textwidth]{ra1.pdf} 
   %  \caption{ Left,  right panels show $\phi$ and $r$ across the bounce for different values of $\beta$, when $p=0.08$, $g_0=1.2$ and $b_g=0.5$.} 
 %\label{fig:r1}
%\end{figure}
%In figure (\ref{fig:r1}) we show the value of $r$ for different parameter values of $p=0.05$, $g_0=1.2$ and $b_g=0.5$ without performing the small $b_g$ approximation.
The tensor to scalar ratio spans several decades $10^{-15}\lesssim r \lesssim 10^{-2}$. %Clearly, it is possible to obtain a large range of tensor-to-scalar ratios by changing the parameters. 
Figure (\ref{fig:g}) depicts the function $g$ for different parameter values that give different $r$. We see that on average a longer bounce gives a smaller enhancement of the scalar spectrum and as a result a larger $r$. The asymmetry factor ensures that the bounce is indeed stable and non-singular.
\begin{figure}[H]
    \centering
    \includegraphics[width=0.45\textwidth]{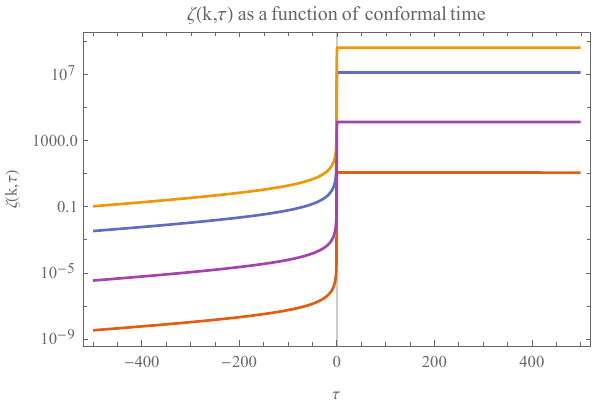} 
    \hspace{0.5cm}
     \includegraphics[width=0.45\textwidth]{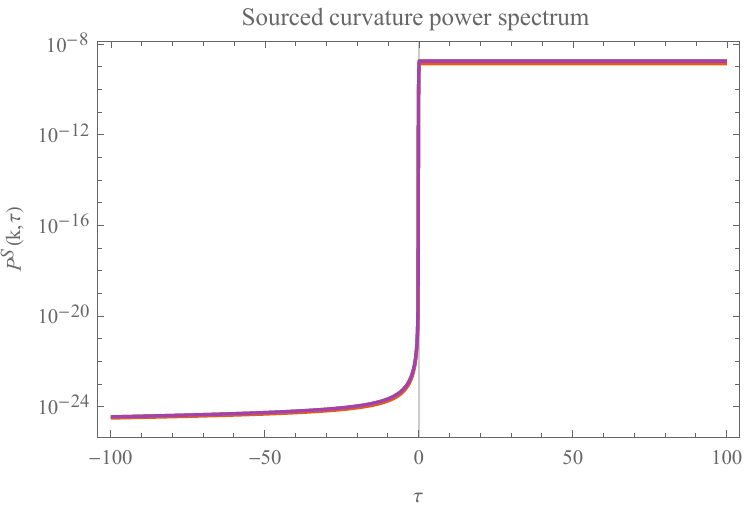} 
     \\
      %\includegraphics[width=0.45\textwidth]{sourcedbbsshort.pdf} 
    %\hspace{0.5cm}
     %\includegraphics[width=0.45\textwidth]{sourced bbs ps short1.pdf} 
       \includegraphics[width=0.45\textwidth]{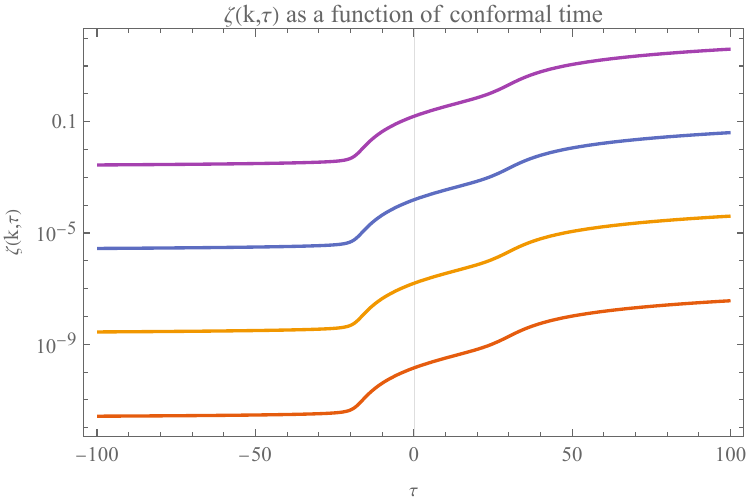} 
    \hspace{0.5cm}
     \includegraphics[width=0.45\textwidth]{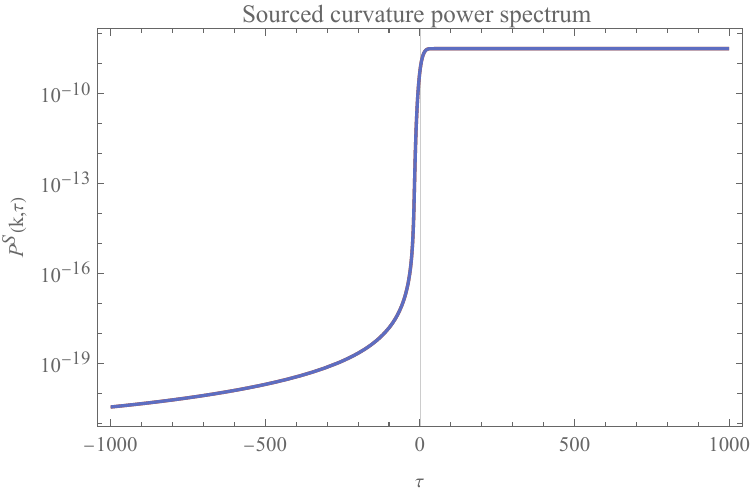} 
     \\
      \includegraphics[width=0.45\textwidth]{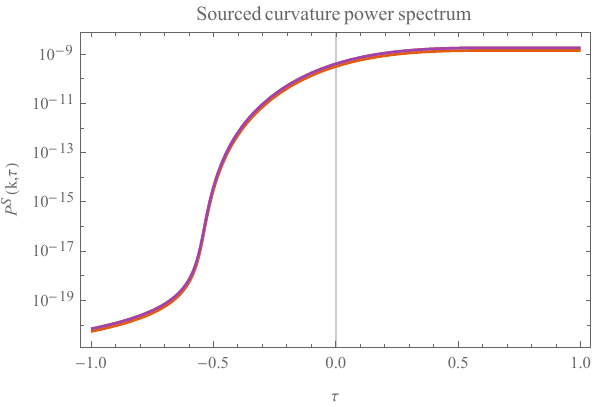} 
    \hspace{0.5cm}
     \includegraphics[width=0.45\textwidth]{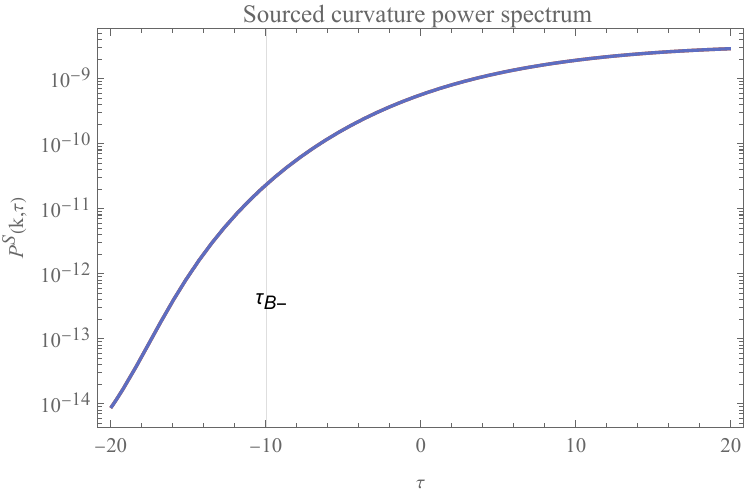} 
     \caption{Left and right panels show Fourier modes of scalar fluctuations and their power spectrum, respectively for different values of parameters (up and down). 
     The amplitude is changing significantly across the bounce and the scale dependence of the power spectrum for super-horizon modes is nearly zero.} 
\label{fig:sourced scalar spectrum}
\end{figure}
We evaluated the sourced spectrum across the bounce numerically. Our numerical results agree with our analytical calculations to a good extent. 
In figure \ref{fig:sourced scalar spectrum}, we have the numerical evaluation of scalar perturbations in a nonsingular bouncing universe of our chosen model with  parameter values 
$V_0 = 10^{-7},g0 = 1.1,\beta = 5,\gamma = 10^{-3},b_V = 5,
b_g = 0.5, p = 0.05 ,q = 0.1$, energy density of gauge fields at $t=0$ at $10^{-8}$ and $\xi$ = 2.4 in the top panel and $V_0 = 10^{-7},g0 = 1.2,\beta = 200,\gamma = 10^{-3},b_V = 0.5,
b_g = 0.28, p = 0.08 ,q = 0.1$, energy density of gauge fields at $t=0$ at $10^{-8}$ and $\xi$ = 2.9. We see that it is possible to match the observed amplitude of scalar perturbations i.e. $A_S = 2.1 \times 10^{-9}$ by choosing appropriate parameters for the theory. In agreement with our theoretical predictions, we see that for $b \sim 0.1$ and large $\beta$ ($\beta=200$ corresponds to $\rho \sim 0.01$) the amplification during the bounce is of order $O(10^2-10^3)$. Furthermore, by adjusting parameters such that the scalar amplitude agrees with the observations, the numerical solution to tensor perturbations (given in the figure \ref{fig:sourced tensor spectrum}) agrees with our theoretical results producing viable values of $r$.
%\begin{figure}[H]
    %\centering
   % \includegraphics[width=0.45\textwidth]{sourced tensor 3.pdf} 
   % \hspace{0.5cm}
     %\includegraphics[width=0.45\textwidth]{sourced tensor ps3.pdf} 
    % \\
 %\caption{Left and right panels show Fourier modes of tensor fluctuations and their power spectrum respectively for $g_0=1.8$.} 
%\label{fig:sourced tensor spectrum2}
%\end{figure}
\section{Fast roll expansion}
\label{sec:fr}
After the bounce, the phase of ghost condensation will stop at $\tau_{B+}$ and the Universe will enter an expanding era. Since the potential of the scalar field $\phi$ is very small in this era, it is dominated by the kinetic term, with an equation of state $w\approx1$. The gauge fields remain subdominant to the scalar field. The source term $J$ in this regime is still given by equation \eqref{eq:sourceb}. However, $a(\tau)$ is not a constant anymore and the source term is decaying as $1/a(\tau)$. 
In this period the equation of motion for cosmological perturbations is given by
\begin{equation}
    v_k''+(k^2+\frac{1}{4\tau^2}) v_k \approx J^S_{\lambda}(k,\tau).
\end{equation}
This equation yields the solution
\begin{equation}
    v_k \simeq e_1(k) \sqrt{(\tau)} J_0(k\tau) + e_2(k) \sqrt{(\tau)} Y_0(k\tau) + \frac{\mathcal{K}}{\tau_{B-}^2}\mathcal{H}_{n}(k\tau),
    \label{eq:kd}
\end{equation}
where $\mathcal{H}_{n}(k\tau)$ is the Struve Hankel function. Matching the solution after the bounce given in equation \eqref{eq:green1} with the above solution at $\tau_{B+}$ and expanding the result at the super horizon limit we obtain that
\begin{equation}
    v_k(k,\tau) \simeq v_{ekp}\cosh(\int_{\tau_{B-}}^{{\tau_{B+}}}\omega(\tau))\times \frac{1}{2}\sqrt{\frac{\tau}{\tau_{B+}}} \log\left(\frac{\tau}{\tau_{B+}}\right).
\end{equation}
The last term is the modification due to kinetic domination and is the same for both the tensor and scalar fluctuations.
While the universe expands
with an equation of state $w = 1$, the scale factor evolves
as  $a \propto \tau^{\frac{1}{2}}$.
This implies that the growing mode of perturbation variable $v_{k}$
evolves proportionally to the scale factor and therefore
the curvature perturbation $\zeta$ will become conserved on
super-Hubble scales after the bounce, thus $\zeta(\tau_{B+}) = \zeta_{kd}$ (i.e. $\zeta$ during kinetic domination). 
Hence, the late-time power spectrum of the curvature fluctuations is 
\begin{equation}
\begin{split}
    P_S(k) = \frac{k^3}{2\pi^2} \bra{\zeta_{k,ekp}}\ket{\zeta_{k,ekp}} F_{\zeta}^2\\
    \end{split}
\end{equation}
Considering what we know about the sourced power spectrum of the contracting universe,
\begin{equation}
\begin{split}
    P_S(k)  = \abs{\hat{f^S}}^2  \frac{225 e^{4\pi \xi}}{32 \pi b^4 \xi^6} \left(\frac{H_{B-}}{M_{pl}}\right)^4 \left(\frac{k}{H_{B-}}\right)^{n_s-1} F_{\zeta}^2.\\
    \label{e:finspec}
    \end{split}
\end{equation}
where 
\begin{equation}
\abs{\hat{f}^S}^2=\int \frac{d^3p}{(2\pi)^\frac{3}{2}}  \abs{\epsilon_i\epsilon_i}^2 \left(p\abs{\vec{k}-\vec{p}}\right)^{2n+1}\\
\end{equation}
is the momentum integral calculated in \cite{r4}. For $n=-2.01$, $f^S=1.68$ and 
\begin{equation}
\begin{split}
    P_S(k)  = 2.8  \frac{225 e^{4\pi \xi}}{32 \pi b^4 \xi^6} \left(\frac{H_{B-}}{M_{pl}}\right)^4 \left(\frac{k}{H_{B-}}\right)^{n_s-1} F_{\zeta}^2.\\
    \label{e:finspec1}
    \end{split}
\end{equation}
and gives the observed power spectrum with $A_S\simeq2.1\times 10^{-9}$(for appropriate values of $H_{B-}$) and $n_s\simeq0.96$, where we have used the same parameter values of the previous section.
\section{Conclusions}
We have examined the evolution of curvature and tensor perturbations from ekpyrotic contraction sourced by a $U(1)$ gauge field across a non-singular bounce. In order to model the non-singular bounce we introduce a Galileon field $\phi$ with a non-canonical kinetic term, which gives rise to the ekpyrotic contraction of the Universe for large and negative values, of $\phi$. As $\phi$ rolls down the potential and becomes small, we enter an era of ghost condensation which gives rise to a non-singular bounce followed by an era of expanding kinetic domination for larger positive values of $\phi$. 

We conclude from our calculations that tensor perturbations
remain unchanged as long as the bounce is short. However, scalar perturbations are amplified as they cross the bouncing regime modifying the tensor to scalar ratio, $r$, and driving it to smaller allowed values. Thus, solving a significant issue of a large $r$ that plagued previous models that assumed that fluctuations remain unchanged across the bounce. With an asymmetric bounce, and choosing appropriate parameters it is possible to obtain a value of $r$ and in particular $r \lesssim 10^{-2}$ which is within the observable limit of future CMB and gravitational wave experiments, and a scalar amplitude and tilt matching current observations. %Assuming a fast roll scenario during the contracting phase, the scale dependence of such a power spectrum is nearly scale-invariant, and with the right parameters of the theory we were able to match the observed CMB scalar spectrum. 
We, therefore, have a viable model of a bounce that explains the observed CMB spectrum and has unique predictions for GW experiments to be launched in the future. This model shows that the measurement of $r$ need not be proof of inflation. An important distinction is that the tensor spectrum of the ``sourced bounce" is chiral and is thus different from the tensor spectrum of inflation. It will be interesting to derive the predictions of our model for laser interferometer experiments such as LIGO and LISA.%Allowing GW experiments to differentiate it from inflation models. 
Recent developments in Observations, background GWs from NANOGrav data, and the blue-tilted spectrum predicted by this data are interesting to study in the context of the blue-tilted vacuum spectrum\cite{Ben-Dayan:2023lwd,Vagnozzi:2023lwo}. 
%Let us stress that the phase of contraction need not be a power law. 
More generally, the contracting phase can be driven by any fast roll potential. The sourced terms can be of any kind, and of course, one can consider various couplings between the scalar field and the would be the source. The Sourced bounce with fast roll contraction is thus a viable Early Universe paradigm of models that is worth pursuing. %can produce a viable CMB spectrum and observable gravitational waves. 
%%%%%%%%%%%%%%%%%%%%%%%%%%%%%%%%%%%%%%%%%%%%%%
\appendix
\section{Solving perturbation equations across bounce}
\label{app:greens}
Fourier modes of perturbation equations during the ekpyrotic regime are denoted by $v_{ekp,s}$, and their detailed derivation can be found in \cite{Artymowski:2020pci,r1}. 
As we approach the bounce, consider modes where $\tau>\tau_B^-$. The equation for Fourier modes in this regime is schematically given by \eqref{eq:vsourcedb}
\begin{equation}
   (\partial^2_{\tau} - \omega^2)v_{b,s} = J,
   \label{eq:vbounce}
\end{equation}
for both tensor and scalar modes, where $J$ is the source term.
We demand continuity of the modes and their derivative at $\tau_B^-$. Laplace transforms are ideally suited for solving inhomogeneous initial value problems like the one at hand. For tractable analytical solutions, we perform a change of variables from $\tau$ to $x=\tau-\tau_{B-}$. Under this transformation the new initial conditions are $v_{b,s}(0)=v_{ekp,s}(\tau_{B-})$ and  $v'_{b,s}(0)=v'_{ekp,s}(\tau_{B-})$.
Laplace transform of \eqref{eq:vbounce} is 
\begin{equation}
   (s^2 - \omega^2)V_{b,s}(s)-V_{b,s}(0)s-V'_{b,s}(0) = L(J)(s)
   \label{eq:vbouncel}
\end{equation}
where $V_{b,s}$ is the Laplace transform of $v_{b,s}$. Solving for $V_{b,s}$
\begin{equation} \label{eq:LT}
    V_{b,s} =  \left(\frac{L(J)(s)}{ (s^2 - \omega^2)} + \frac{ V_{b,s}(0)s+V'_{b,s}(0)}{(s^2 - \omega^2)} \right)
\end{equation}
The inverse Laplace transform of the above equation is 
\begin{equation}
   v_{b,s} = L^{-1} \left(\frac{L(J)(s)}{ (s^2 - \omega^2)}\right) + v_{b,s}(0) \cosh(\omega x) + \frac{v'_{b,s}(0)}{\omega} \sinh(\omega x).
   \label{eq:vbouncef}
\end{equation}
Let us calculate $L(J)(s)$. For tensors, from equation \eqref{sourceb}, the source term during bounce is $J^T_{\lambda,bounce}=\frac{\mathcal{K}}{\tau^2_{B-}}$,
where $\mathcal{K}$ is the time-independent momentum integral. Thus, the source is approximately constant in time across the bounce and
\begin{equation}
%\begin{split}
   LJ(s) = L(\frac{\mathcal{K}}{\tau^2_{B-}}) = \frac{\mathcal{K}}{\tau^2_{B-}} \frac{1}{s}.
   % \end{split}
\end{equation}
Observe that 
\begin{equation}
   % \begin{split}
        \frac{L(J)(s)}{ (s^2 - \omega^2)} =  \left( \frac{\mathcal{K}}{\tau^2_{B-}} \frac{1}{s
        } \frac{1}{s^2 - \omega^2} \right) \\
        = \frac{\mathcal{K}}{\tau^2_{B-}} \frac{1}{ \omega^2} \left( \frac{s}{s^2 - \omega^2} - \frac{1}{s} \right) 
   % \end{split}
\end{equation}
and the inverse Laplace transform yields
\begin{equation}
\begin{split}
   L^{-1} & \left(\frac{L(J)(s)}{ (s^2 - \omega^2)}\right) =  \frac{\mathcal{K}}{\tau^2_{B-}} \frac{2\sinh^2\left( \omega \frac{x}{2}\right)}{\omega^2}.
   \label{eq:invla}
   \end{split}
\end{equation}
Combining equations \eqref{eq:invla}, \eqref{eq:vbouncef} and changing the variable back to $\tau$ we obtain that
\begin{equation}
\begin{split}
v_{b,s} &= v_{ekp,s}(\tau_{B-}) \cosh(\omega(\tau-\tau_{B-}))+ \frac{1}{\omega} v_{ekp,s}'(\tau_{B-}) \sinh(\omega(\tau-\tau_{B-})) \\
&  \;\;\;\;\;\;\;\;\;\;\;\;\;\;\;\;\;\;+\frac{\mathcal{K}}{\tau^2_{B-}} \frac{2\sinh^2\left( \omega (\tau-\tau_{B-})\right)}{\omega^2}
\label{eq:vi2} 
\end{split}
\end{equation}
%\subsection{Scalar perturbations}

For scalars, \eqref{eq:LT} is still the solution, but with a different source term and $\omega_S$ is specified by \eqref{eq:omegas}. During the bounce $I$ is nearly constant, and $f_{\alpha} \simeq \gamma \frac{\phi'^3}{2\mathcal{H}}$, then we can write \eqref{eq:c3} as
\begin{equation}
    C_{bounce}(\tau) = \frac{1}{z(\tau)}\left(\frac{6 \phi''}{\gamma \phi'^4}\right).
\end{equation}
We also know that during bounce phase $z\simeq 3 a\beta/\gamma \phi' $, thus
\begin{equation}
    C_{bounce}(\tau) \simeq  \frac{2\phi''}{ \beta \phi'^3}  \simeq  \frac{4\tau}{\beta T^2 (\phi'_B)^2 e^{-2\tau^2/T^2}}.
\end{equation}
While considering scalar perturbations the source term $J^S_{\lambda,bounce}=\frac{\mathcal{K}}{\tau^4_{B-}}  C_{bounce}(\tau)$, thus 
\begin{equation}
    J^S_{\lambda,bounce}(k,\tau) = \frac{\mathcal{K}}{\tau^2_{B-}}   \frac{4\tau}{\beta T^2 (\phi'_B)^2 e^{-2\tau^2/T^2}}
\end{equation}
and
%\begin{equation}
   % LJ(s) = \frac{\mathcal{K}}{8\tau^2_{B-}} \frac{4}{\beta T^2 (\phi'_B)^2} \left(4 + 
   %e^{\left(-\frac{1}{8} s^2 T^2\right)} \sqrt{2 \pi} s i \;\; erfc\left((s T)/(2 \sqrt{2} i)\right)\right). 
%\end{equation}
\bea
    LJ(s) &=& \frac{\mathcal{K}}{\tau^2_{B-}} \frac{4}{\beta T^2 (\phi'_B)^2} \frac{T^2}{16} \left[-4 + 
   e^{- s^2 T^2/8} \sqrt{2 \pi} s T\left( -i + erfi\left((s T)/(2 \sqrt{2})\right)\right)\right]\cr 
%\end{equation}
%\begin{equation}
    %LJ(s) = 
    &=&\frac{\mathcal{K}}{4\tau^2_{B-}\beta\phi^{'2}_B} \left[-4 + 
   e^{- s^2 T^2/8} \sqrt{2 \pi} s T\left( -i + erfi\left((s T)/(2 \sqrt{2})\right)\right)\right]. 
\eea

Where $erfi$ is the complex error function. 
Assuming that $\omega_S = \frac{z''}{z}$ is constant, sourced perturbations across bounce is given by
\begin{equation}
   v_{b,s} = L^{-1} \left(\frac{L(J)(s)}{ (s^2 - \omega_S^2)}\right) + v_{b,s}(0) \cosh(\omega_S x) + \frac{v'_{b,s}(0)}{\omega_S} \sinh(\omega_S x).
   \label{eq:vbouncef1}
\end{equation}
Note that
%\begin{eqnarray}
%\begin{split}
  % L^{-1}  \left(\frac{L(J)(s)}{ (s^2 - \omega_S^2)}\right)
   %&=&   \frac{4}{\beta T^2 (\phi'_B)^2} \frac{\mathcal{K}}{8\tau^2_{B-}} \Bigg[\sinh(\omega_S \tau)\left(\frac{4}{\omega_S} + \sqrt{\pi/2} e^{-\omega_S^2T^2/8}erf(T\omega_S/2\sqrt{2})\right)+\cr
  % &+&e^{-\omega_S^2T^2/8} \left( erf\left(\frac{4\tau+T^2\omega_S}{2\sqrt{2}T}\right) e^{-\omega_S \tau}+erf\left(\frac{4\tau-T^2\omega_S}{2\sqrt{2}T}\right) e^{\omega_S \tau}\right)\Bigg].\cr
  % &\simeq&  \frac{2}{\beta T^2 (\phi'_B)^2} \frac{\mathcal{K}}{\tau^2_{B-}} \frac{\sinh(\omega_S \tau)}{\omega_S}.
   \label{eq:invl}
  % \end{split}
%\end{eqnarray}
\be
%\begin{split}
   L^{-1}  \left(\frac{L(J)(s)}{ (s^2 - \omega_S^2)}\right)
   \simeq\frac{\mathcal{K}}{\tau^2_{B-}\beta\phi^{'2}_B}\frac{\sinh(\omega_S \tau)}{\omega_S}
   \ee
   \label{eq:vi1}
Hence 
%\begin{equation}
%begin{split}
%v_{b,s} &= v_{ekp,s}(\tau_{B-}) \cosh(\omega_S(\tau-\tau_{B-}))+ \frac{1}{\omega_S} v_{ekp,s}'(\tau_{B-}) \sinh(\omega_S(\tau-\tau_{B-})) \\
%&  \;\;\;\;\;\;\;\;\;\;\;\;\;\;\;\;\;\;+ \frac{2}{\beta T^2 (\phi'_B)^2} \frac{\mathcal{K}}{\tau^2_{B-}} \frac{\sinh( \omega_S \tau)}{\omega_S}
%\label{eq:vi1} 
%\end{split}
%\end{equation}

\be
v_{b,s} = v_{ekp,s}(\tau_{B-}) \cosh(\omega_S(\tau-\tau_{B-}))+ \frac{1}{\omega_S} v_{ekp,s}'(\tau_{B-}) \sinh(\omega_S(\tau-\tau_{B-}))+ \frac{\mathcal{K}}{\tau^2_{B-}\beta\phi^{'2}_B}\frac{\sinh(\omega_S (\tau-\tau_{B-}))}{\omega_S}
\ee

Now instead of constant $\omega_S$, we use the WKB approximation such that $\omega_S(\tau-\tau_{B-})$ is replaced by the integral $\int^{\tau}_{\tau_{B-}}\omega_S(\tau') d\tau'$.
\section{Second order equations from action} \label{app:second order equations}
In order to obtain second-order equations of motion, first we vary actions \eqref{eq:ac3rd} and \eqref{eq:ac2nd} with respect to $\zeta$. Variation of terms that do not contain the inverse Laplacian term is straightforward, we hereby present the result of this variation with respect to $\zeta$,
\begin{equation}
\begin{split}
    &\zeta'' z + 2  z' \zeta' + c_s^2  z \partial^2 \zeta + a^2 \partial_i\zeta \partial^i \zeta+\frac{a^2}{2f_{\alpha}\mathcal{H}}\left(-16\partial^2 \zeta+3\partial^2(\partial^2\zeta) \right) \partial_t \zeta \\
    & \frac{a^2}{2f_{\alpha}\mathcal{\mathcal{\mathcal{H}}}}\left(-16\partial_i \zeta \partial_t(\partial^i\zeta)+6\partial_i(\partial^2\zeta) \partial_t(\partial_i\zeta)-4\partial_t\partial_i\zeta \partial_t\partial^i\zeta +(\partial_t\zeta)^2(9f_{\alpha}\mathcal{H}-9m_{\alpha}\right)\\
    &+ \frac{a^2}{2f_{\alpha}\mathcal{H}} \left( \partial_t \partial_i \zeta \partial_t \zeta(1-3f_{\alpha}\mathcal{H}-8)+\partial^2 \zeta \zeta'' (4+6f_{\alpha}\mathcal{H}\kappa-1) \right)\\
    &+ \frac{a^2}{2f_{\alpha}\mathcal{H}}\left((36-9f_{\alpha}\mathcal{H}\kappa(4+3f_{\alpha}\mathcal{H}\kappa)-2\lambda_3+2\lambda 5 \mathcal{H}) \zeta'\zeta''+4\zeta( f_{\alpha}\mathcal{H}\partial^2\zeta +9(f_{\alpha}\mathcal{H}-m_{\alpha}) \zeta'' ) \right)\\
\end{split}
\end{equation}
$\lambda_3$ and $\lambda_5$ are described by equations \eqref{eq:lambda3} and \eqref{eq:lambda5}. 
As for the rest of the terms, they are obtained by varying the equation \eqref{eq:invlap} given below with respect to $\zeta$,
\begin{equation}
\begin{split}
& \int dt d^3x  a^2\left( \zeta' \partial_i \zeta \partial^i  \partial^{-2}\left( \frac{1}{\mathcal{H} f_{\alpha}}  \partial^2 \zeta-3\kappa \zeta'\right)\right) + a^2\left( \left( \partial^i  \partial^{-2}\left( \frac{1}{\mathcal{H} f_{\alpha}}  \partial^2 \zeta-3\kappa \zeta'\right) \right)^2 \partial^2\zeta\right)\\
&+\frac{1}{2} a^2\left(\zeta \left( \partial^2  \partial^{-2}\left( \frac{1}{\mathcal{H} f_{\alpha}}  \partial^2 \zeta-3\kappa \zeta'\right) \right)^2 - \zeta \left( \partial_i \partial_j  \partial^{-2}\left( \frac{1}{\mathcal{H} f_{\alpha}}  \partial^2 \zeta-3\kappa \zeta'\right) \right)^2\right) \\
    &- 2a^2 \mathcal{H} f_{\alpha} \alpha \left(  \partial_i \zeta \partial^i  \partial^{-2}\left( \frac{1}{\mathcal{H}  f_{\alpha}}  \partial^2 \zeta-3\kappa \zeta'\right)\right) - \alpha \left(\partial_i \partial_j  \partial^{-2}\left( \frac{1}{\mathcal{H}  f_{\alpha}}  \partial^2 \zeta-3\kappa \zeta'\right) \right)^2. \\
    \label{eq:invlap}
\end{split}
\end{equation}
As an example for variation of the terms with inverse Laplacian operator gather all the terms with '$\partial_i \partial_j$' spatial derivatives, varying them in a straightforward manner we obtain
\begin{equation}
\begin{split}
    &\fdv{\left(\int \frac{1}{2} (\zeta+\frac{1}{\mathcal{H}f_{\alpha}} \zeta') \left( \partial_i \partial_j \partial^{-2}\left( \frac{1}{\mathcal{H}  f_{\alpha}}  \partial^2 \zeta-3\left(\frac{m}{f_{\alpha}^2}-1 \right) \zeta' \right)\right)^2\right)}{\zeta} = \\
    & \frac{1}{2} \left( \partial_i \partial_j \partial^{-2}\left( \frac{1}{\mathcal{H}  f_{\alpha}}  \partial^2 \zeta-3\left(\frac{m}{f_{\alpha}^2}-1 \right) \zeta' \right)\right)^2 \\
    &+ \frac{1}{\mathcal{H} f_{\alpha}} \partial_\tau\left( \partial_i \partial_j \partial^{-2}\left( \frac{1}{\mathcal{H}  f_{\alpha}}  \partial^2 \zeta-3\left(\frac{m}{f_{\alpha}^2}-1 \right) \zeta' \right)\right)^2\\
   &+  \frac{1}{2} (\zeta+\frac{1}{\mathcal{H}f_{\alpha}} \zeta') \fdv{ \left( \partial_i \partial_j \partial^{-2}\left( \frac{1}{\mathcal{H}  f_{\alpha}}  \partial^2 \zeta-3\left(\frac{m}{f_{\alpha}^2}-1 \right) \zeta' \right)\right)^2 }{\zeta} .
    \end{split}
    \label{eq:varex}
\end{equation}
In order to understand the last part of the equation \eqref{eq:varex} notice that 
\begin{equation}
    \grad^{-2} f(\zeta) = \int e^{-u \grad^2} f(\zeta) du
\end{equation}
Since $\partial_i, \partial_j$ commute with $\grad$ we can write 
\begin{equation}
  \partial_i \partial_j  \grad^{-2} f(\zeta) =    \int e^{-u \grad^2} \partial_i \partial_j f(\zeta) du = \grad^{-2} \partial_i \partial_j f(\zeta)
\end{equation}
Now if we assume E to be the Euler-Lagrange operator namely
\begin{equation}
   E= \partial_{\mu} \partial_{\zeta_{\mu}}-\partial_{\zeta}
\end{equation}
then
\begin{equation}
\begin{split}
    E e^{-u \grad^2} &= \comm{E}{e^{-u \grad^2}} + e^{-u \grad^2}E \\
    &= \comm{E}{\grad^2}  e^{-u \grad^2} + e^{-u \grad^2}E 
    \end{split}
\end{equation}
Thus
\begin{equation}
   E \grad^{-2} f(\zeta)  = \grad^{-2} E f(\zeta) + \comm{E}{\grad^2}  \grad^{-2}.
\end{equation}
All second-order terms obtained this way upon expanding over $\zeta=\zeta_1+\zeta_2$, are functions of $\zeta_1,\zeta_1',\partial \zeta_1$ and will lead to a blue tilted spectrum that is exponentially suppressed compared to the perturbations sourced by gauge fields. 
\subsection{Second order equations}
\label{app:source ekp}
We use the expansion of $\zeta$ given by $\zeta=\zeta_1+\zeta_2$ and collect all the second-order terms which will result in the equation
\begin{equation}
\begin{split}
    &\zeta_2'' z + 2  z' \zeta_2' + c_s^2  z \partial^2 \zeta_2 + a^2 \partial_i\zeta_1 \partial^i \zeta_1+\frac{a^2}{2f_{\alpha}\mathcal{H}}\left(-16\partial^2 \zeta_1+3\partial^2(\partial^2\zeta_1) \right) \partial_t \zeta_1 \\
    & +\frac{a^2}{2f_{\alpha}\mathcal{H}}\left(-16\partial_i \zeta_1 \partial_t(\partial^i\zeta_1)+6\partial_i(\partial^2\zeta_1) \partial_t(\partial_i\zeta_1)-4\partial_t\partial_i\zeta_1 \partial_t\partial^i\zeta_1 +(\partial_t\zeta_1)^2(9f_{\alpha}\mathcal{H}-9m\right)\\
    &+ \frac{a^2}{2f_{\alpha}\mathcal{H}} \left( \partial_t \partial_i \zeta_1 \partial_t \zeta_1(\lambda_4-3f_{\alpha}H-8)+\partial^2 \zeta_1 \zeta_1'' (4+6f_{\alpha}\mathcal{H}\kappa-\lambda_4) \right)\\
    &+ \frac{a^2}{2f_{\alpha}\mathcal{H}}\left((36-9f_{\alpha}\mathcal{H}\kappa(4+3f_{\alpha}\mathcal{H}\kappa)-2\lambda_3+2\lambda 5 \mathcal{H}) \zeta_1'\zeta_1''+4\zeta_1( f_{\alpha}\mathcal{H}\partial^2\zeta_1 +9(f_{\alpha}\mathcal{H}-m) \zeta_1'' ) \right)\\
    &+C=0\\
\end{split}
\end{equation}
where $C$ is the sum of all the terms that contain the inverse Laplacian operator. 
Terms containing 
$\zeta_1$ (including terms in the sum $C$) contribute to a blue-tilted sourced spectrum much like the terms containing $\delta \varphi_1$ in \cite{r4}, and can promptly be ignored following a similar logic. Only the second-order perturbations and the sourced term need to be considered, hence the equation for sourced perturbations are 
\begin{equation}
    \zeta_2'' z + 2  z' \zeta_2' + c_s^2 k^2  \zeta_2  = J_s(\tau,k)
    \label{sms}
\end{equation}
Sourced term $J_s(\tau,k)$ is calculated by varying the term in the action containing the gauge fields with respect to $\zeta$. 
\section{Source term during ekpyrosis} \label{app:C}
During the regime of ekpyrosis $\epsilon \gg 1$ and $\epsilon \approx \frac{\dot{\phi}^2}{2 H^2}$, $f_{\alpha}=1$. Substituting the above in equation \eqref{eq:sourcece} 
\begin{eqnarray}
%\begin{split}
    J_s &=& \frac{1}{2 (2\pi)^\frac{3}{2}}\int d^3p \Sigma_{\lambda}\hat{\mathcal{P_{\lambda}}}\left( \dfrac{d}{d\tau} \left(-\frac{I^2}{f_{\alpha}\mathcal{H}} \left( \dfrac{d A_{\lambda}}{d\tau} \right)^2 \right) -  I^2 \left( \dfrac{d A_{\lambda}}{d\tau} \right)^2\right)\cr
 %   \end{split}
  %  \label{eq:sourcece}
%\end{equation}
%\begin{equation}
%\begin{split}
      &=&  -\frac{1}{2 (2\pi)^\frac{3}{2}}\int d^3p \Sigma_{\lambda}  \hat{\mathcal{P_{\lambda}}} \left(   -\frac{\mathcal{H}'}{\mathcal{H}^2}I^2\left(\dfrac{d A_{\lambda}}{d\tau}\right)^2+ \dfrac{d I^2}{d\tau} \frac{1}{\mathcal{H}}  \left(\dfrac{d A_{\lambda}}{d\tau}\right)^2 + \frac{I^2}{\mathcal{H}}\dfrac{d}{d\tau}\left(\dfrac{d A_{\lambda}}{d\tau}\right)^2 +   I^2 \left(\dfrac{d A_{\lambda}}{d\tau}\right)^2\right)  \cr
      &=& - \frac{1}{2 (2\pi)^\frac{3}{2}}\int d^3p\Sigma_{\lambda}  \hat{\mathcal{P_{\lambda}}} \left(\epsilon\left(\dfrac{d A_{\lambda}}{d\tau}\right)^2\left( I^2+ \dfrac{d I^2}{d\tau} \frac{1}{\mathcal{\epsilon H}}   \right) + \frac{I^2}{\mathcal{H}}\dfrac{d}{d\tau}\left(\dfrac{d A_{\lambda}}{d\tau}\right)^2\right).
 %    \end{split}
     \label{eq:cinb}
\end{eqnarray}
Note that from $A_{\lambda}=\frac{\tilde{A}_{\lambda}}{I}$ and equation \eqref{eq:guage}
\begin{equation}
    I^2\dfrac{d}{d\tau} \left(\dfrac{d A_{\lambda}}{d\tau}\right)^2 =  -2\dfrac{d I^2}{d\tau} \left(\dfrac{d A_{\lambda}}{d\tau}\right)^2 + k^2 A_{\lambda}\dfrac{d A_{\lambda}}{d\tau}.
    \label{eq:newe}
\end{equation}
For $k\tau\ll1$, the second term in equation \eqref{eq:newe} can be safely ignored similar to terms containing  magnetic fields. Equation \eqref{eq:cinb} can thus be rewritten as

\be
J_s=- \frac{1}{2 (2\pi)^\frac{3}{2}}\int d^3p\Sigma_{\lambda}\hat{\mathcal{P_{\lambda}}}\epsilon\left(\dfrac{d A_{\lambda}}{d\tau}\right)^2\left( I^2- \dfrac{d I^2}{d\tau} \frac{1}{\mathcal{\epsilon H}}   \right) 
\ee

\section{Expansion of \texorpdfstring{ $\dot{\phi}$}{dp}}
A series expansion of equation \eqref{eq:dbp} is obtained by perturbatively solving for $\dot{\phi}$ and putting it back in equation \eqref{eq:dbp}. We hereby write down the exact expressions for coefficients up to third-order expansion. 
\label{app:dphi}
\begin{equation}
     c_1 = \frac{4 b_g^2 (-1+g_0)g_0}{\left(6b g_0-\left(\frac{1}{b_g}\right)^{-\frac{1}{1+b_g}}(1+2 g_0) \right) \beta p}
     \label{eq:c1}
\end{equation}
\begin{equation}
    c_2 = \frac{4(-1+b_g)b_g^{2}g_0\frac{1}{p}^{\frac{3}{2}}\frac{-1+g_0}{\beta}^{\frac{3}{2}}}{3\sqrt{3}\left(6b_g g_0-\left(\frac{1}{b_g}\right)^{-\frac{1}{1+b_g}}(1+2 g_0) \right) }
    \label{eq:c2}
\end{equation}
and
\begin{equation}
\begin{split}
   c_3 &= \frac{1}{9 \left(6b_g g_0 \frac{1}{b_g}^{\frac{1}{1+b_g}}-(1+b_g)(1+2 g_0)\right)^3 p^2 \beta^2} 2 ((1/b_g)^{(-2 + 1/(1 + b_g))}) ((-1 + g_0)^2) g_0 \times \\
   & (36 (1/b_g)^{(-((2 b_g)/(1 + b_g)))} g_0^2 + 
 36 (-5 + b) (1/b_g)^{(-3 + 2/(1 + b_g))} g_0^2 +(1 + 2 g_0)^2 - 5 b_g (1 + 2 g_0)^2 \\
 &- 5 b_g^3 (1 + 2 g_0)^2+b^4 (1 + 2 g_0)^2 - 12 (b_g + 2 b g_0)^2 \\
 &-6 (1/b_g)^{(-1 + 1/(
  1 + b_g))} (1 + b_g) g_0 (2 + 4 g1 + b^2 (2 + 4 g_0) - 3 b_g (5 + 7 g_0)))
   \end{split}
\end{equation}

\label{beb} 
\bibliographystyle{JHEP}
\bibliography{main}

\end{document}